%% file: main.tex
\documentclass[fleqn,10pt]{wlscirep}
\usepackage[utf8]{inputenc}
\usepackage[T1]{fontenc}
\usepackage{pdfpages}

\usepackage{tikz}
\usepackage[percent]{overpic}
\usetikzlibrary{arrows, positioning, backgrounds, shadows}

\usepackage{wrapfig}
\usepackage{fancyhdr}

\tikzstyle{materia}=[draw, fill=blue!20, text width=6.0em, text centered, minimum height=1.5em,drop shadow]
\tikzstyle{practica} = [materia, text width=24em, minimum width=24em, minimum height=2em, rounded corners, drop shadow]
\tikzstyle{practicao} = [materia, text width=11em, minimum width=11em, minimum height=2em, rounded corners, drop shadow]
\tikzstyle{texto} = [above, text height=2em, text centered]
\tikzstyle{linepart} = [draw, thick, color=black!50, -latex', dashed]
\tikzstyle{line} = [draw, thick, color=black!50, -latex']
\tikzstyle{ur}=[draw, text centered, minimum height=0.01em, text width = 3em]
\tikzstyle{uro}=[draw, text centered, minimum height=0.01em, text width = 6em]
 
\newcommand{\practica}[2]{node (p#1) [practica]
  {\scriptsize\textit{#2}}}

\newcommand{\background}[5]{%
  \begin{pgfonlayer}{background}
    \path (#1.west |- #2.north)+(-1,0.4) node (a1) {};
    \path (#3.east |- #4.south)+(+0.4,-0.2) node (a2) {};
    \path[fill=yellow!20,rounded corners, draw=black!50, dashed]
      (a1) rectangle (a2);
    \path (a1.east |- a1.south)+(0.8,-0.3) node (u1)[texto]
      {\scriptsize\textit{#5}};
  \end{pgfonlayer}}

\newcommand{\transreceptoro}[3]{%
  \path [linepart] (#1.east) -- node [above]
    {\scriptsize #2} (#3);}

\AtBeginShipout{%
  \ifnum\value{page}=2%
    \lfoot{*Made by 3doptix optical system design. https://3doptix.com.}%
    \AtBeginShipoutNext{%
      \AtBeginShipoutNext{%
        \lfoot{}%
      }%
    }
  \fi%
}

\title{SSR-PR: Single-shot Super-Resolution Phase Retrieval based two prior calibration tests}

\author[1,*]{P\'eter Kocsis}
\author[1]{Igor Shevkunov}
\author[1]{Vladimir Katkovnik}
\author[2]{Heikki Rekola}
\author[1]{Karen Egiazarian}
\affil[1]{Tampere University, Faculty of Information Technology and Communication Sciences, Tampere, FI-33101, Finland}
\affil[2]{Institute of Photonics, University of Eastern Finland, P.O. Box 111, FI-80101 Joensuu, Finland}

\affil[*]{peter.kocsis@tuni.fi}

\begin{abstract}
We propose a novel approach and algorithm based on two preliminary tests of the optical system elements to enhance the super-resolved complex-valued imaging. The approach is developed for inverse phase imaging in a single-shot lensless optical setup. Imaging is based on wavefront modulation by a single binary phase mask. The preliminary tests compensate errors in the optical system and correct a carrying wavefront, reducing the gap between real-life experiments and computational modeling, which improve imaging significantly both qualitatively and quantitatively. These two tests are performed for observation of the laser beam and phase mask along, and might be considered as a preliminary system calibration. The corrected carrying wavefront is embedded into the proposed iterative Single-shot Super-Resolution Phase Retrieval (SSR-PR) algorithm. Improved initial diffraction pattern upsampling, and a combination of sparse and deep learning based filters achieves the super-resolved reconstructions. Simulations and physical experiments demonstrate the high-quality super-resolution phase imaging. In the simulations, we showed that the SSR-PR algorithm corrects the errors of the proposed optical system and reconstructs phase details 4$\times$ smaller than the sensor pixel size. In physical experiment 2 $\mu$m thick lines of USAF phase-target were resolved, which is almost 2$\times$ smaller than the sensor pixel size and corresponds to the smallest resolvable group of used test target.  For phase bio-imaging, we provide Buccal Epithelial Cells reconstructed in computational super-resolution and the quality was of the same level as a  digital holographic system with 40$\times$ magnification objective. Furthermore, the single-shot advantage provides the possibility to record dynamic scenes, where the framerate is limited only by the used camera. We provide an amplitude-phase video clip of a moving alive single-celled eukaryote.
\end{abstract}

\begin{document}

\flushbottom
\maketitle
\thispagestyle{empty}

\input{Sections/1_Introduction}

\input{Sections/2_Problem_formation}

\input{Sections/3_Enhancements}

\input{Sections/4_Simulations}

\input{Sections/5_Physical_experiments}

\input{Sections/6_Conclusion}

\bibliography{main}

\end{document}

%% file: Sections/1_Introduction.tex
\section*{Introduction}

Complex-valued object imaging has been long studied in a wide range of tasks over the past decades and leads to significant developments in quantitative phase microscopy. The methods used for complex imaging %in phase microscopy 
are usually based on two main techniques, which are holography\cite{gabor1948new} or phase retrieval\cite{fienup1982phase}.
The history of phase retrieval goes back more than 50 years ago to Sayre's observations of the Bragg diffraction\cite{sayre1952some}. He captured the diffraction pattern of a coherently illuminated sample and recognized that an adequate high sampling rate will result in a unique real-space image of the sample. This idea led to the early phase retrieval method of coherent diffraction imaging\cite{miao1999extending} (CDI). 
Contrary to holography where the phase reconstruction is made from a hologram, which is the interference pattern between the object and reference beams, in phase retrieval only a single beam is used. This single beam is disturbed by an object and the intensity of the diffracted wavefront is captured by a sensor as a diffraction pattern.
The fact that only the intensity of the light radiation can be captured while the phase is lost in all observations, results in an ill-posed problem called \textit{phase problem}. The Gerchberg-Saxton algorithm\cite{gerchberg1972practical} provides a good tool to solve this problem by iterative forward- and backward propagation between the object and the sensor planes. In each iteration, the amplitude of the wavefront is updated by the captured images, hence the errors from the ill-conditioning are reduced.
% A far-field imaging system is assumed in which the propagation was defined by Fourier transform. 
The technique is depending on two sets of images, as one set on each plane. This dependence can be bypassed by a prior knowledge about \textit{support constraints} (e.g. non-negativity or known apodization diameter) on the planes. Applying these constraints in every iteration, the reconstruction error decreases, therefore this generalization is called error-reduction algorithm\cite{fienup1982phase}. Afterward, Fienup proposed an additional time-domain correction step to improve the convergence rate, and led to the well-known Hybrid Input-Output\cite{fienup2003phase} (HIO) algorithm. 

In the present day, the literature provides several number of methods to solve the phase problem. As A.Fannjiang and T.Strohmer state\cite{fannjiang2020numerics} the natural way to overcome the ill-posedness is by reducing the number of unknown parameters. One of the most common approach is using the above mentioned \textit{support constraints} on the signal \cite{kang2021single, dong2018single}. Another more recent and expansively studied solution is sparsity\cite{shechtman2015phase}.
It includes a special prior knowledge based on the assumption that the observed object $x$ has some known sparse representation, as $x=\Psi \alpha$. The representation matrix $\Psi$ is called sparsity basis, while $\alpha$ stands for the sparse vector. In the most basic case the object is composed of few point sources, so $\Psi$ will be an identity matrix.
% in which several small patches of the image assumed to have similar features. These patches located in different parts of the image and taken together they give a sparse representation.

In most approaches, the key to solve the phase problem is to capture the propagated object wavefront on several decorrelated diffraction patterns (e.g., \cite{shevkunov2014comparison}). The decorrelation can be generated by lateral shearing\cite{de2020extreme}, using tilt diffraction model\cite{hu2020coherent}, displacing the sensor plane along the optical axis\cite{ling2020three}, or ptychography\cite{zhou2020low,zheng2021concept,yan2020ptychographic}. Another possibility is to extend the system with more illumination wavelengths. The advantage of the multi-wavelength method is the capability to work with spatially incoherent light\cite{tahara2020multiwavelength}, and the color imaging possibility\cite{marien2020color}.
Special programmable devices can be also used to obtain decorrelated images, such as Spatial Light Modulator (SLM)\cite{wang2020high} or Digital Micromirror Device (DMD)\cite{deng2018characterization}.

Most developed techniques are using far-field approach, since it requires a much simpler propagation model, which can be defined by Fourier transform only.
However, with the trend to miniaturization in computer technology, the far-field techniques are not adequate anymore because they are not implementable in a compact system. The near-field techniques provide a good solution, but they require changes in the propagation model, since the diffraction pattern differs significantly. 
It is due to the fact that in far-field approach the spherical waves are flattened out far from the source, so they are handled as planar wavefronts. However, in near-field approach, this assumption is not valid anymore, and the wavefronts have to be treated as spherical.
The Angular Spectrum (AS) method can solve the wave equation exactly for near-field diffraction, since the formula is extracted from the Rayleigh-Sommerfeld diffraction theory without approximation. 

A significant advantage of the phase retrieval method is the lensless imaging capability, which further increases the compactness. 
% The compactness can be further increased by omitting the lenses, which is a significant advantage of the phase retrieval methods. 
The eschew of lens makes the system light and cost-effective while free from lens aberration and provides larger field of view\cite{jiang2020wide}.
% The drawback of the lensless system is that the captured pattern is a merged overlapping image of several diffraction orders, which raises difficulties in reconstruction. 
The literature provides alternatives to imitate the lenses in imaging with special modulation masks, so-called Fresnel lenses\cite{wu2020single,mirirostami2021extended}. Although the Fresnel lenses are also sensitive to chromatic aberrations when used with broadband sources and distort the formed images. 
Another lensless approach - which we followed - is to use wavefront modulation with modulation mask\cite{boominathan2020phlatcam, kocsis2020single, zhang2016phase} or diffuser\cite{antipa2018diffusercam}. They distort the illumination carrying wavefront and provide a coded pattern on the sensor with known modulation. This coded pattern is a wide-spread diffraction pattern, which from more data can be collected. Although using this technique, the mask also generates corruption in phase reconstruction. It increases the loss from ill-conditioning, therefore several observations with different masks are typically used for better convergence\cite{katkovnik2017computational}.
The literature provides techniques to reconstruct complex-valued images from a single coded pattern using wavefront modulation\cite{horisaki2017subpixel,antipa2018diffusercam}, however, it is a very challenging task due to the corruption generated by the modulation.
% so the results are suffering from the noise generated by the spreading. 

In our previous paper\cite{kocsis2020single} design and modelling of single-exposure lensless system is proposed with wavefront modulation by random binary phase mask. The successful super-resolved reconstruction is relying on \textit{support constraint} on the object plane with known diameter of the laser beam and a sparsity-based block-matching 3D (BM3D)\cite{makinen2020collaborative} filter. These methods are embedded into a modified iterative Super-Resolution Sparse Phase Amplitude (SR-SPAR)\cite{kocsis2019single} algorithm. 
% Since only one observation is used, the method can be applied for 3D imaging dynamic samples. 
% Although these results are standing out in the field of single-shot lensless phase imaging\cite{zhou2018single,chang2020single,hassan2020lensless} 
The super-resolved reconstructions are demonstrated in simulations and physical experiments, however, the algorithm has difficulty to reconstruct small phase values and provides a generally noisy output.

The main contribution of the current paper is a new model approach and a new algorithm based on preliminary calibration of the optical system.
To overcome the problems of the previous solution, the problem formulation has been rethought. We came out with a new approach, resulting in a Single-shot Super-Resolution Phase Retrieval (SSR-PR) method. The SSR-PR provides qualitative and quantitative phase imaging for a wider range of object's phase values in high resolution and with sufficiently higher noise suppression. In SSR-PR, the object wavefront at the object plane is separated into 2 parts: studied object and a novel introduced error compensated carrying wavefront. The formation of the phase retrieval problem with this new approach is presented in Sec.\ref{sec:problem}. The calibration includes two prior experiments which are made without the studied object: the laser beam and the mask-diffracted laser beam. Using together, we can estimate a compensation on the object plane (Sec.\ref{sec:cpo}). The results of these two prior experiments are processed with improved initial upsampling (Sec.\ref{sec:upsampling}) by the novel SSR-PR algorithm (Sec.\ref{sec:sr-spar}).
We provide simulation experiments (Sec.\ref{sec:simul}) to demonstrate the high-quality super-resolution capability by reconstructing complex objects with a resolution of 4$\times$ less than the pixel size of the camera.
The SSR-PR method has significant advantages since it is able to correct the errors caused by the manufacturing tolerance of the optical elements, such as modulation mask phase-shift difference, dull pixel edges, or even the unknown curved wavefront from the illumination (Sec.\ref{sec:errorcomp}). Furthermore, it corrects the computational modeling of the optical image formation, which results in increased quality of the phase imaging.
As for demonstration, three physical experiments are performed with different test targets. The first target was a calibrated United States Air Force (USAF) resolution phase test chart. It demonstrates the super-resolution capability by resolving the smallest details of 2 $\mu$m, almost 2$\times$ smaller than the sensor pixel size. The second target was a static biological sample of Buccal Epithelial cells. On the one hand this experiment provides evidence on biomedical applications, on the other hand it shows that the SSR-PR method can properly reconstruct objects with maximum phase-shift over 2$\pi$ (wrapped phase objects). The third target was a dynamic biological sample as a moving single-celled eukaryote. We recorded 287 diffraction patterns over 10 seconds and post-processed them by SSR-PR method, resulting in a super-resolved video of the moving biological specimen.
This successful video reconstruction is a breakthrough, since as the best of our knowledge it is the first super-resolved video reconstructed by phase retrieval with an adequately high frame rate.

%% file: Sections/2_Problem_formation.tex
\section{Problem formation}
\label{sec:problem}

\begin{figure}[b!]
    \centering
    \begin{overpic}[trim={0cm 0cm 0cm 1cm}, clip, width=.9\textwidth]{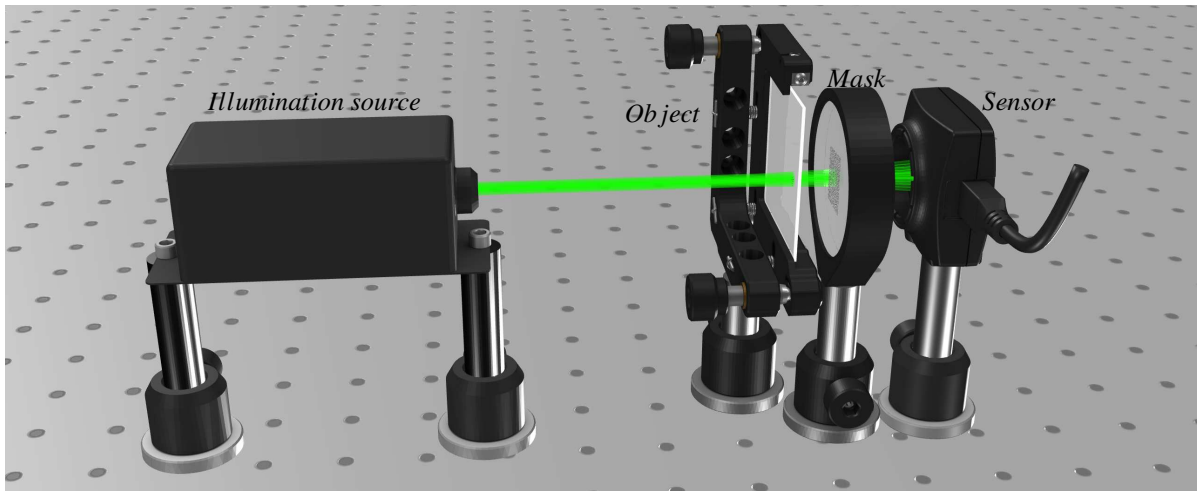}
    \put(17,32){\color{black}{$Illumination~source$}}
    \put(52,31){\color{black}{$Object$}}
    \put(69,34){\color{black}{$Mask$}}
    \put(82,32){\color{black}{$Sensor$}}
    \end{overpic}
    \caption{Sketch of the SSR-PR optical system
    : A coherent illumination source (laser), an object to be tested, a binary phase-mask, and a sensor.*}
    \label{fig:scheme}
\end{figure}

The typical lensless phase retrieval scheme includes a coherent light source (e.g., a laser beam) and an object to be reconstructed. The carrying wavefront $u_{b,0}$ is diffracted by the complex object $u_{o}$ then the intensity pattern of the diffracted wavefront on the sensor plane can be written as
\begin{equation}
    \label{eq:form_base}
    z=|P_d \left \{u_{o} \cdot u_{b,0}\right \}|^2.
\end{equation}
 Here $u_{o}$ and $u_{b,0}$ are 2D functions and the operator $P_{d}$ stands for the free space forward propagation on the distance $d$. Typically, the carrying wavefront is assumed to be well-known and flat ($u_{b,0}=1$). 
%  This assumption might be inaccurate, therefore using $u_{b,0}$ we can add a corrected estimation from the illumination wavefront.
%  Since the illumination wavefront $u_{b,0}$ is unknown, it is usually taken out of the equation, so it can be handled as all-ones matrix with zero phase shift.
%  Since the phase is lost a single pattern is usually not enough to resolve the phase problem, therefore the iterative reconstruction becomes stagnant. 
The phase problem consists of the fact that only the intensity of the light radiation ($z$) can be captured while the phase is lost. It is an ill-posed problem, therefore, traditionally quantitative phase reconstruction is not possible from a single diffraction pattern without a prior knowledge of \textit{support constraint}. As it was developed in our previous approach\cite{kocsis2020single}, we apply a random binary modulation phase mask $M$ in the system as shown in Fig. \ref{fig:scheme}. Taking this modulation into account Eq.~(\ref{eq:form_base}) can be rewritten as
% we use this approach
\begin{equation}
    \label{eq:form_prev}
    z = \left | P_{d_{2}}\left \{ M\cdot P_{d_{1}}\left \{u_{o} \cdot u_{b,0} \right \}\right \} \right |^{2},
\end{equation}
where $d_{1}$ and $d_{2}$ correspond to the object-mask and mask-sensor distances. The beam is diffracted by the mask, propagates forward, and spread across the sensor resulting in a coded intensity  pattern. This coded pattern covers larger area of the sensor, therefore more data can be collected. More data and the prior known mask pattern can offer enough information to improve the complex-valued object reconstruction.
The wavefront propagation is modelled by the Rayleigh-Sommerfeld theory, in which the angular spectrum (AS) method\cite{goodman2005introduction} is defined as:
\begin{equation}
    \label{eq:prop}
    u(x,y,d)=\mathfrak{F}^{-1} \left \{ H(f_{x},f_{y},d)\cdot\mathfrak{F} \left \{ u(x,y,0) \right \} \right \},
\end{equation}
\begin{equation}
    \label{eq:trfun}
    H\left (f_{x},f_{y},d \right)=\begin{cases} exp\left [ i\frac{2\pi }{\lambda }d\sqrt{1-\lambda ^{2}\left ( f_{x}^{2} + f_{y}^{2}\right )} \right ], & f_{x}^{2} + f_{y}^{2}\leq \frac{1}{\lambda ^{2}},\\ 0,  & otherwise.\\\end{cases}
\end{equation}
The method determines the free space propagation of $u(x,y,0)$ in distance $d$ resulting in $u(x,y,d)$, where $\mathfrak{F}$ and $\mathfrak{F}^{-1}$ operators stand for the Fourier and inverse Fourier transforms. The AS operator $H(f_{x},f_{y},d)$ is defined by the distance $d$, the spatial frequencies $f_{x}$,$f_{y}$ and the wavelength $\lambda$. 

The coded diffraction pattern has to be decoded with the known phase mask and distances to achieve decent reconstructions. As we showed in our previous paper\cite{kocsis2020single}, if we know our parameters ($\lambda$, $d$, $M$, etc.) perfectly the ill-conditioning can be resolved by the modified SR-SPAR algorithm and the phase reconstructed unequivocally. The achieved resolution is limited only by the transfer function (Eq.~(\ref{eq:trfun})). However, in physical experiments we can use approximated values of our parameters at most. Due to the error of approximations, we had difficulties to reconstruct small phase values and the retrieved object was disordered by the errors of the theoretically precise inputs.

The proposed SSR-PR method follows a novel approach by prior calibration of the optical system (Fig. \ref{fig:scheme}) to improve the fidelity of the computational model of our physical system. 
% to make the approximated parameters closer to the real ones. 
The following algorithm has been found to compensate efficiently the discrepancy between the reality of our optics and the models used for wavefront reconstruction. In Eq.~(\ref{eq:form_prev}) \ $u_{b,0}$ is replaced by a complex-valued compensated carrying wavefront on the object plane as
\begin{equation}
    \label{eq:cpo}
    \widehat{u_{b}} = u_{b,0} \cdot CPO. 
\end{equation}
Here, $u_{b,0}$ and the Compensation by Prior Observation ($CPO$) are 2D functions calculated by using the resulting diffraction patterns of two prior experiments as presented in Sec.\ref{sec:cpo}. 
% These patterns include the errors of the optical elements and the experiments help us estimate the needed error compensation of the experimental setup in relation to the ideal computational model. 
% These errors can be extracted and a complex-valued compensation ($CPO$) can be calculated.
These diffraction patterns allow us to estimate the proper behavior of our computation model and a complex-valued compensation ($CPO$) can be calculated. 
% From these diffraction patterns, we can extract the erroneous behaviour of our computation model and a complex-valued compensation ($CPO$) can be calculated. 
The compensation is able to correct the errors caused by the optical elements, improve the carrying wavefront by the missing phase, and increase the image formation model correspondence with reality.
% The obtained compensated wavefront is used in the modified SR-SPAR algorithm, which could be implemented into any phase retrieval system with light modulation.
Moreover, we improved the initial diffraction pattern upsampling for computational super-resolution reconstructions of SSR-PR. Several interpolation methods were analyzed and the most adequate technique was selected as stairstep interpolation with Lanczos-3 kernel. It already proved its effectiveness in previous researches connected to remote sensing\cite{madhukar2013lanczos} and medical imaging\cite{moraes2020medical}.

%  \textcolor{red}{Say from the very beginning what is a contribution of this paper !!!!!!!}

%  \textcolor{red}{
% Main comments:\\(1) The following heuristic has been found to compensate efficiently  the discrepancy between the reality of our optics and the models used for wavefront reconstruction. In Eq.(2) $u_{b}$ is replaced by the product shown in (5), where cpo is a 2D function calculated by using the results of the following two experiments....\\
% (2) Some problems with notation for $u_{b}$ in different use\\
% (3) Ill-posedness is not corrected in this approach. It is not proved;\\
% (4) Move section 2.2 before 2.1;\\
% (5) The text require an attention\\
% Should be discussed sentence-by-sentence}

%% file: Sections/3_Enhancements.tex
\section{SSR-PR Algorithm}
\label{sec:novel}

In this section, we describe the novel enhancements of the SSR-PR method. In Sec.\ref{sec:upsampling} the enhancements of the initial diffraction pattern upsampling are presented to achieve super-resolution reconstruction. Using the upsampled diffraction patterns we introduce a novel approach in Sec.\ref{sec:cpo} to give an approximation of the compensated carrying wavefront. This compensated wavefront and the filters initiated in Sec. \ref{sec:filters} are embedded into the SSR-PR method (Sec.\ref{sec:sr-spar}).

\subsection{Diffraction pattern upsampling for super-resolution}
\label{sec:upsampling}

\begin{figure}[b!]
    \centering
    \begin{overpic}[trim={5cm 1.8cm 3cm 2.8cm}, clip, width=\textwidth]{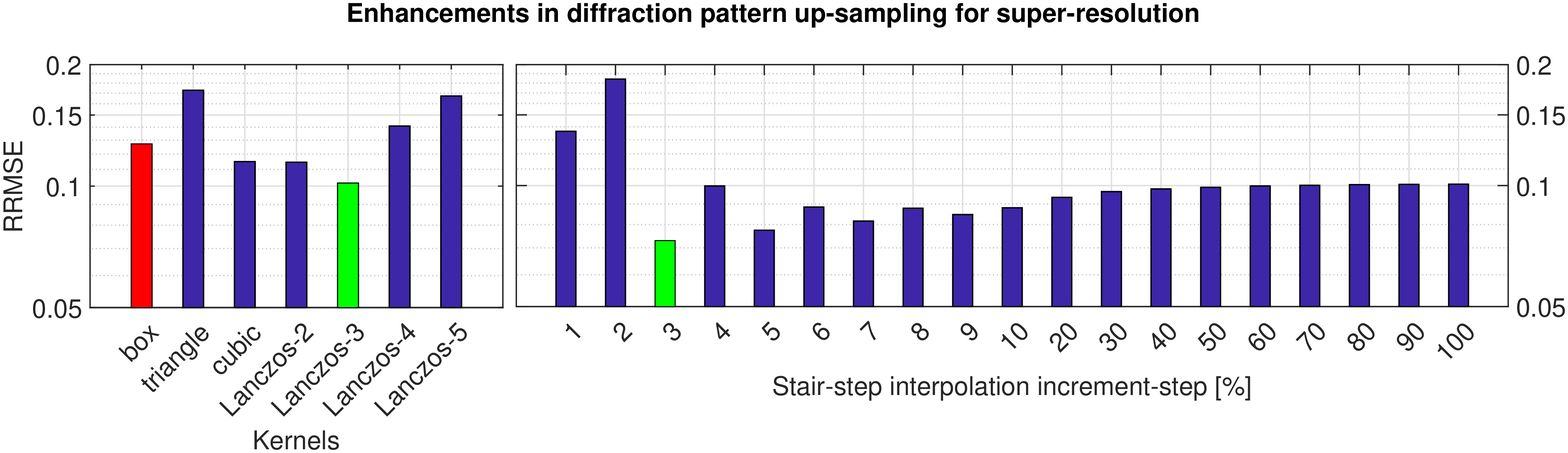}
    \end{overpic}
    \caption{RRMSE of phase reconstructions by (a) different upsampling kernels and (b) stairstep interpolation of Lanczos-3 kernel with different increment-steps.}
    \label{fig:kernels}
\end{figure}

The computational wavefront propagation requires discretization of the continuous diffraction patterns by a computational pixel size of $\Delta_{c} \times \Delta_{c}$. Physically the sampling is performed by the sensor, therefore the resolution is limited by the sensor pixel size of $\Delta_{s} \times \Delta_{s}$. This means that traditionally the highest resolution is achieved with $\Delta_{c}=\Delta_{s}$, which is also called \textit{pixel-wise resolution}. We are talking about computational \textit{super-resolution} if the smallest resolved details are smaller than the sensor pixel size $\Delta_{s}$.
A general approach for \textit{super-resolution} imaging is using multiple slightly shifted observations. If the captured patterns are merged properly, the result is an upsampled, super-resolved diffraction pattern, from which super-resolved image of the object is retrievable\cite{gerchberg1974super}. The required subpixel resolution can be also achieved by special upsampling techniques developed for single image processing\cite{rivenson2010single}.

In this paper, we followed a different approach based on the modeling of optical image formation and registration. We found that using upsampled diffraction patterns as input for the iterative method the resolution increases.
In our preceding paper\cite{kocsis2020single} we demonstrated that using this technique with ideal \textit{support constraint} and BM3D filter, the resolution is limited only by the AS transfer function (Eq.~\ref{eq:trfun}). 
% If $\Delta_{c}/\lambda<0.7$, the values at the edges of the AS transfer function become zeros, hence the high frequencies will be eliminated. 
This limitation occurs if the computational pixel size is too small, and as a result the high frequencies will be eliminated.
Due to the elimination, the problem will be assumed more ill-posed and the reconstructions will be not acceptable. In our system ($\lambda=532$ nm) this limit occurs if $\Delta_{c}<376$ nm. This limit corresponds to a super-resolution factor of $r_{s}=\Delta_{s}/\Delta_{c}=9.17$, which means that the diffraction patterns are initially upsampled by 9$\times$. In nonideal case, the reconstructions are disturbed by the errors of the optical system and these errors are accumulated in the upsampling. We found that taking $r_{s}>4$ the resolution will be not improved observably, but the calculation time will significantly increase.

Previously in SR-SPAR, the diffraction patterns were upsampled by a common box interpolation kernel and super-resolution factor of $r_{s}$.
However, we recognized that the resolution of the reconstructions is depending on the interpolation method of the upsampling. For SSR-PR, we have tested the most commonly used interpolation kernels as follows:
\begin{enumerate}
    \item \textit{box kernel}: nearest-neighbor interpolation with pixel value duplication
    \item \textit{triangle kernel}: bilinear interpolation with distance-weighted averaging in the $2\times2$ nearest neighborhood
    \item \textit{cubic kernel}: bicubic interpolation with distance-weighted averaging in the $4\times4$ nearest neighborhood
    \item \textit{Lanczos-2 kernel}: Lanczos interpolation based on the 3-lobed Lanczos window function
    \item \textit{Lanczos-3 kernel}: Lanczos interpolation based on the 5-lobed Lanczos window function
\end{enumerate}
The interpolation kernels were used to upsample the input diffraction patterns of the SSR-PR algorithm (Sec.\ref{sec:sr-spar}) and the Relative Root-Mean Square errors (RRMSE) of the reconstructions are shown in Figure \ref{fig:kernels}. The RRMSEs are calculated by
\begin{equation}
\label{eq:rmse1}
    RRMSE=\frac{\left \| \varphi_{o}-\widehat{\varphi_{o}} \right \|_{F}}{\left \| \varphi_{o} \right \|_{F}},
\end{equation}
where $\left \| \cdot \right \|_{F}$ stands for the Frobenius norm, while $\varphi_o$ and $\widehat{\varphi_o}$ are the phases of the original and the reconstructed object.

As it is shown, using Lanczos-3 kernel, the RRMSE decreases by 20\% compared to the previously used box kernel. Then we applied stairstep interpolation with the selected Lanczos-3 kernel. In this method, we continuously resize the pattern with the same increment to 400\% ($r_{s}=4$). If the ratio between the maximum size-increase of 300\% and the increment-step is not an integer, then the last increment-step is taken to get exactly the required upsampling size. The lowest RRMSE is achieved by using increment-step of 3\% with $RRMSE<0.1$, which count as a successful reconstruction.

\subsection{CPO calculation}
\label{sec:cpo}

\begin{figure}[b!]
\vspace{-20pt}
    \centering
    \input{Sections/tikz/CPO_flowchart}
    \caption{Flowchart to calculate $CPO$-compansated carrying wavefront.}
    \label{fig:flowchart}
\end{figure}
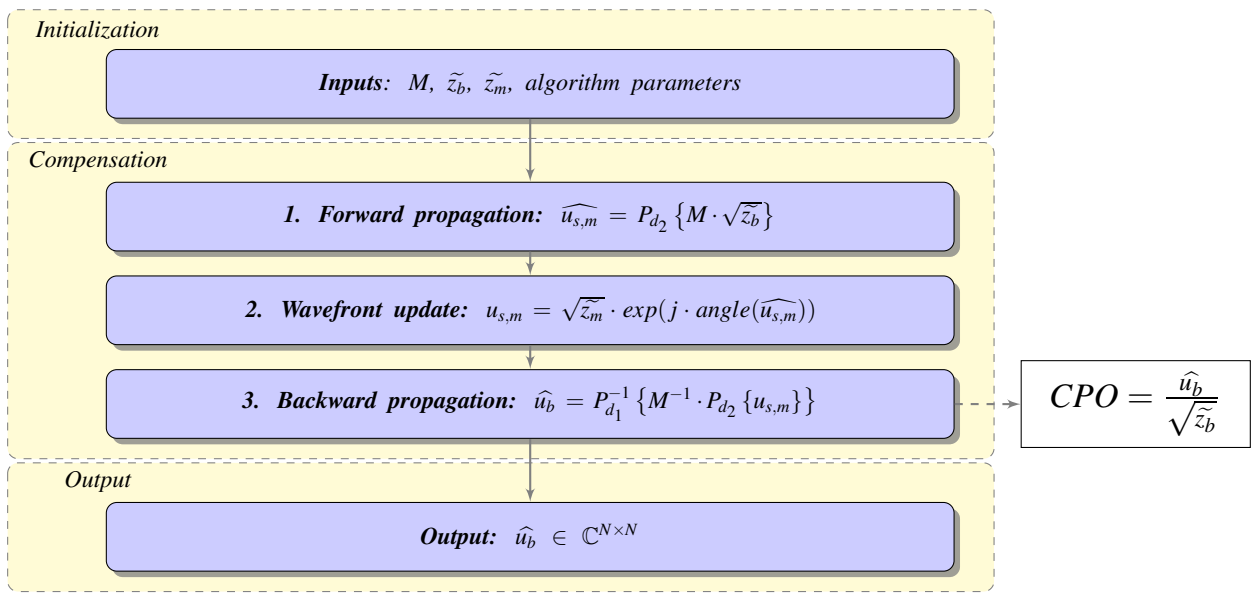

The proper reconstructions require precisely known prior knowledge about the optical system to solve the phase problem. however, discrepancies could appear between the 
computation model and reality. These unnoted discrepancies in the illumination, mask parameters, propagation modeling and distances are resulting in deficient parameter approximations and inaccurate image formation.
% expected and real values. These unnoted discrepancies in the illumination, mask parameters, and propagation distances are resulting in imperfect approximations of these parameters. The errors of these approximations and manufacturing inaccuracies are accumulated, resulting in an erroneous output.
Using \textit{support constraint} and \textit{sparsity}-based filter can moderately reduce this noise, but small phase values are lost\cite{kocsis2020single} for small details.

The SSR-PR method is aiming to expand our prior knowledge, hence providing a calibration to the optical system. We define the compensation $CPO$ which has a ternal function: give a proper estimation for the phase part of the carrying wavefront, compensate the corruptions appearing due to the errors of modulation mask and the distances, and correct the approximations in wavefront propagation.
% by better approximations by error compensation. We define the compensation $CPO$ which has a dual function: give a proper estimation for the phase part of the carrying wavefront and compensate the corruption which appear due to the errors of modulation mask and the distances. 
Two prior experiments were initiated as follows:
In the first experiment, the mask and the object is omitted and we capture the intensity pattern of illumination source (\textit{beam}) only as $z_{b}$. This pattern can be used to give a better approximation of the carrying wavefront as $u_{b,0}=\sqrt{z_{b}}$, instead of the general assumption of $u_{b,0}=1$. We assume that coherent plane waves are generated by the laser with the same wavefront on all planes. In the second experiment, the mask is placed back and the diffraction pattern is captured as $z_{m}$. Using the upsampling techniques on the diffraction patterns $z$, $z_{b}$, and $z_{m}$ as presented in Sec.\ref{sec:upsampling}, they will result in $\widetilde{z}$, $\widetilde{z_{b}}$, and $\widetilde{z_{m}}$.

We are assuming that our laser has the same amplitude on all planes, therefore the carrying wavefront on the mask plane is taken as $\sqrt{z_{b}}$. Since the phase is unknown, this wavefront is just an approximation of the real carrying wavefront, which we aim to correct by $CPO$.
% Two prior experiments were initiated, as stated above (Sec. \ref{sec:problem}): the intensity of the illumination beam as $z_{b}$ and a mask-diffracted coded intensity pattern as $z_{m}$. Since we are assuming that our laser has the same amplitude on all planes, the illumination wavefront on the mask plane is taken as $\sqrt{z_{b}}$. Since the phase is unknown, this wavefront is just an approximation of the real illumination wavefront, which we aim to correct by $CPO$.
The mask $M$ is added to this upsampled carrying wavefront on the mask plane ($M\cdot \sqrt{\widetilde{z_{b}}}$) and they are propagated forward to sensor plane. The amplitude is replaced by $\sqrt{\widetilde{z_{m}}}$ and the updated wavefront is propagated back to the object plane with mask subtraction, resulting in $\widehat{u_{b}}$, from which $CPO$ can be calculated as
\begin{equation}
    \label{eq:cpoMain}
    CPO=\frac{\widehat{u_{b}}}{\sqrt{\widetilde{z_{b}}}}
\end{equation}
The flowchart of this \textit{Compensation} calculation is shown in Fig.~\ref{fig:flowchart}.
however, as we can see from this equation, $\widehat{u_{b}}$ already contains the compensation, therefore calculating $CPO$ is only referential.

\subsection{Enhanced filters}
\label{sec:filters}

The modulation mask scatters the beam, so the high intensities will be scattered and spread across the sensor. Due to this scattering and wide spreading, a larger area of the sensor is in use. Therefore, much more information can be collected, but also more noise is generated and accumulated.
The \textit{sparse}-based BM3D filter already proved its effectiveness for several optical phase recovering problems\cite{katkovnik2012high,katkovnik2013sparse}, and we also used it successfully to filter the reconstructed wavefront\cite{kocsis2020single}. however, occasionally the similar patches of the modulation mask can result in correlated noises, which are corrupting the sparse representation.
To overcome this problem, in SSR-PR the latest version of BM3D\cite{makinen2020collaborative} is used, which focuses on the collaborative filtering of correlated noise.

% To further increase the effectiveness of the noise filtering,
however, BM3D might still fail occasionally, so a state-of-the-art deep learning based, plug and play image restoration (IR) filter was added with Dilated-Residual U-Net (DRUNet) Deep Denoiser Prior\cite{zhang2021plug}. It combines U-net\cite{ronneberger2015u} and ResNet\cite{he2016deep} effectively to create a better precondition for sparse filtering. 
The filter was trained for plug and play IR applications on a large dataset containing almost 9000 images. Furthermore, during the training, different noise levels have been applied, to be able to filter images in wider noise level range.
The experiments showed that using DRUNet before BM3D the RRMSE of the reconstructions decreases more than 20\%.
% 400 Berkeley segmentation dataset (BSD) images\cite{chen2016trainable}, 4744 Waterloo Exploration Database images\cite{ma2016waterloo}, 900 DIV2K dataset images\cite{agustsson2017ntire}, and 2750 Flick2K dataset images\cite{lim2017enhanced}. 
% Furthermore during the training, the noise level $\sigma$ was randomly chosen from [0,50], to be able to filter images in wider noise level range.

Previously apodization with zero values was used to eliminate the appearing noise in the area outside of the \textit{support constraint}. The diameter of the \textit{support constraint} was corresponding to the size of the illumination beam.
In the SSR-PR, the wavefront on the object plane $u_{o}\cdot \widehat{u_{b}}$ is suppressed (divided) by $\widehat{u_{b}}$. The remaining noises outside of the \textit{support constraint} are apodized by assuming a plane wavefront with zero phase shift. This means that these values are replaced by ones.

\subsection{SSR-PR iterations}
\label{sec:sr-spar}

\begin{figure}[t!]
\vspace{-20pt}
    \centering
    \input{Sections/tikz/SRSPAR_flowchart}
    \caption{Flowchart of the SSR-PR method.}
    \label{fig:flowchart2}
\end{figure}
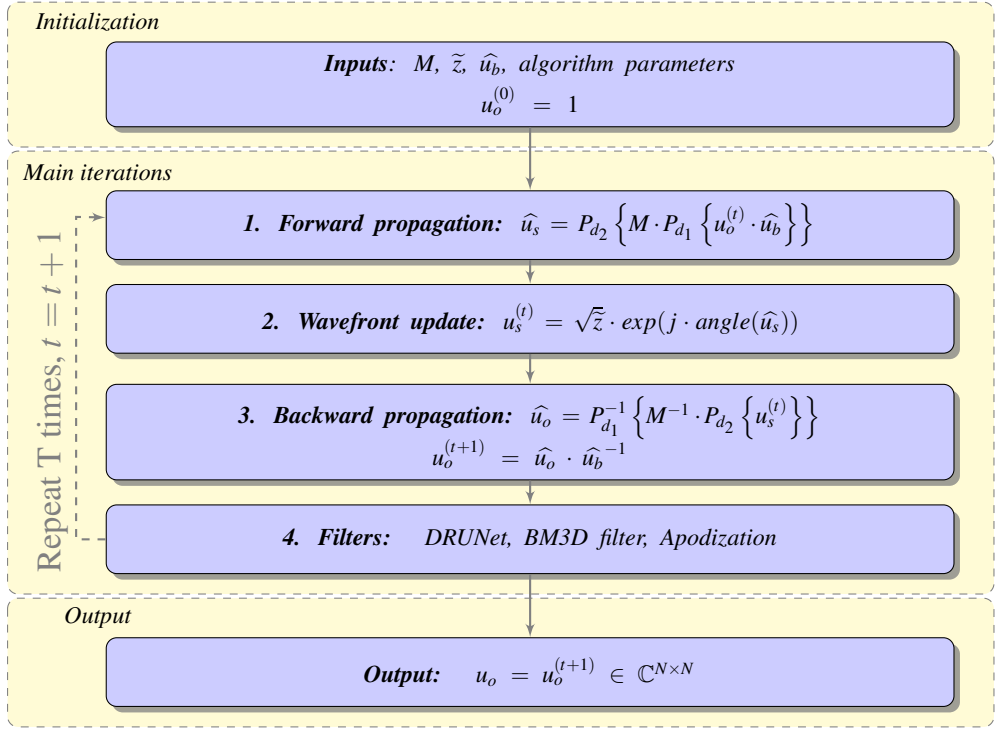

The flowchart of the SSR-PR method is shown in Figure \ref{fig:flowchart2}. 
% In SR-SPAR, the initialization consisted of a back-propagation only as
% \begin{equation}
%     \label{eq:previnit}
%     u_{o}^{0} = P_{d_{1}}^{-1}\left \{ M^{-1}\cdot P_{d_{2}}\left \{ \sqrt{\widetilde{z}} \right \}\right \}.
% \end{equation}
% This initial backward propagation is replaced by calculating the compensated illumination wavefront $\widehat{u_{b}}$ and the initial object $u_{o}$ is taken as all-ones matrix ($u_{o}^{0} = 1$). 
The \textit{initialization} consists of the calculation of the compensated carrying wavefront $\widehat{u_{b}}$ and the initiation of the object as $u_{o}^{(0)} = 1$.
Their sizes are corresponding to the upsampled diffraction pattern of $\widetilde{z}$. The algorithm parameters include the distances ($d_{1}$, $d_{2}$), sensor and mask properties, mask position, wavelength, and filtering properties. The first step of the \textit{Main Reconstruction} is a \textbf{forward propagation} of the complex wavefront $u_{o} \cdot \widehat{u_{b}}$, where $\widehat{u_{b}}$ contains the amplitude of the carrying wavefront and the compensation $CPO$, as in Eq. \ref{eq:cpo}. The second step is a \textbf{wavefront update} with the upsampled amplitude $\sqrt{\widetilde{z}}$ derived from the captured intensity of the mask diffracted object observation. 
The third step starts with \textbf{backward propagation} of the updated wavefront, then divided by the compensated carrying wavefront $\widehat{u_{b}}$. To avoid the division with zero values, we use regularization\cite{maiden2017further} on $\widehat{u_{b}}$ with the weight $\alpha$ as
\begin{equation}
    \label{eq:regularization}
    u_{b}'=\frac{u_{b}^{*}}{\left ( 1-\alpha \right ) \left | u_{b} \right |^{2}+\alpha\left | u_{b} \right |_{max}^{2}},
\end{equation}
where $u^{*}$ stands for the conjugate of $u$.
In the fourth step, the obtained object wavefront's amplitude and phase are separately filtered, then merged into a complex wavefront and apodized. The iteration is run until a criteria is met (e.g., number of iterations, RRMSE).

%% file: Sections/tikz/CPO_flowchart.tex
\begin{tikzpicture}[scale=1.3,transform shape]
  
  % Initiate
  \path \practica{1}{\textbf{Inputs}: \textit{$M$, $\widetilde{z_{b}}$, $\widetilde{z_{m}}$, algorithm parameters}};
  
  % Calculate cpo
  \path (p1.south)+(-0.0,-1.0) \practica{2}{\textbf{1. Forward propagation: $\widehat{u_{s,m}} = P_{d_{2}}\left \{ M\cdot \sqrt{\widetilde{z_{b}}} \right \}$}};
  \path (p2.south)+(-0.0,-0.6) \practica{3}{\textbf{2. Wavefront update: $u_{s,m}=\sqrt{\widetilde{z_{m}}} \cdot exp(j \cdot angle(\widehat{u_{s,m}}))$}};
  \path (p3.south)+(-0.0,-0.6) \practica{4}{\textbf{3. Backward propagation: $\widehat{u_{b}} = P_{d_{1}}^{-1}\left \{ M^{-1}\cdot P_{d_{2}}\left \{ u_{s,m} \right \}\right \}$}};
  \path (p4.east)+(+1.85,0) node (ur1)[uro] {$CPO=\frac{\widehat{u_{b}}}{\sqrt{\widetilde{z_{b}}}}$};
  
  % Output
  \path (p4.south)+(-0.0,-1.0) \practica{5}{\textbf{Output: $\widehat{u_{b}}\in\mathbb{C}^{N \times N}$}};

  % Draw arrows between elements
  \path [line] (p1.south) -- +(+0.0,-0.0) -- +(+0.0,-0.0)
    -- node [above, midway] {} (p2);
  \path [line] (p2.south) -- +(+0.0,-0.0) -- +(+0.0,-0.0)
    -- node [above, midway] {} (p3);
  \path [line] (p3.south) -- +(+0.0,-0.0) -- +(+0.0,-0.0)
    -- node [above, midway] {} (p4);
  \path [line] (p4.south) -- +(+0.0,-0.0) -- +(+0.0,-0.0)
    -- node [above, midway] {} (p5);
  \transreceptoro{p4}{}{ur1}
   
  \background{p1}{p1}{p1}{p1}{Initialization}
  \background{p2}{p2}{p4}{p4}{Compensation}
  \background{p5}{p5}{p5}{p5}{Output}
  
\end{tikzpicture}

%% file: Sections/tikz/SRSPAR_flowchart.tex
\begin{tikzpicture}[scale=1.3,transform shape]

  % Initialization
  \path \practica{1}{\textbf{Inputs}: \textit{$M$, $\widetilde{z}$, $\widehat{u_{b}}$, algorithm parameters\\$u_{o}^{(0)}=1$}};
  
  % Main iteration
  \path (p1.south)+(-0.0,-1.0) \practica{2}{\textbf{1. Forward propagation: $\widehat{u_{s}} = P_{d_{2}}\left \{ M\cdot P_{d_{1}}\left \{u_{o}^{(t)} \cdot \widehat{u_{b}} \right \}\right \}$}};
  \path (p2.south)+(-0.0,-0.6) \practica{3}{\textbf{2. Wavefront update: $u_{s}^{(t)}=\sqrt{\widetilde{z}} \cdot exp(j \cdot angle(\widehat{u_{s}}))$}};
  \path (p3.south)+(-0.0,-0.8) \practica{4}{\textbf{3. Backward propagation: $\widehat{u_{o}} = P_{d_{1}}^{-1}\left \{ M^{-1}\cdot P_{d_{2}}\left \{ u_{s}^{(t)} \right \}\right \}$\\ $u_{o}^{(t+1)}=\widehat{u_{o}} \cdot \widehat{u_{b}}^{-1}$}};
  \path (p4.south)+(-0.0,-0.6) \practica{5}{\textbf{4. Filters: } DRUNet, BM3D filter, Apodization};

  % Output
  \path (p5.south)+(-0.0,-1.0) \practica{6}{\textbf{Output: } $u_{o} = u_{o}^{(t+1)}\in\mathbb{C}^{N \times N}$};
  
  % Draw arrows between elements
  \path [line] (p1.south) -- +(+0.0,-0.0) -- +(+0.0,-0.0)
    -- node [above, midway] {} (p2);
  \path [line] (p2.south) -- +(+0.0,-0.0) -- +(+0.0,-0.0)
    -- node [above, midway] {} (p3);
  \path [line] (p3.south) -- +(+0.0,-0.0) -- +(+0.0,-0.0)
    -- node [above, midway] {} (p4);
  \path [line] (p4.south) -- +(+0.0,-0.0) -- +(+0.0,-0.0)
    -- node [above, midway] {} (p5);
  \path [line] (p5.south) -- +(+0.0,-0.0) -- +(+0.0,-0.0)
    -- node [above, midway] {} (p6);
  \path [linepart] (p5.west) -- +(-0.3,-0.0) -- +(-0.3,+3.3)
    -- node [below, pos=-0.8] {\rotatebox{90}{Repeat T times, $t=t+1$}} (p2);
   
  \background{p1}{p1}{p1}{p1}{Initialization}
  \background{p2}{p2}{p5}{p5}{Main iterations}
  \background{p6}{p6}{p6}{p6}{Output}
  
\end{tikzpicture}

%% file: Sections/4_Simulations.tex
\section{Simulations}
\label{sec:simul}

\begin{figure}[t!]
  \centering
    \setlength{\fboxsep}{0pt}%
    \setlength{\fboxrule}{0pt}%
  \fbox{\begin{overpic}[height=3cm]{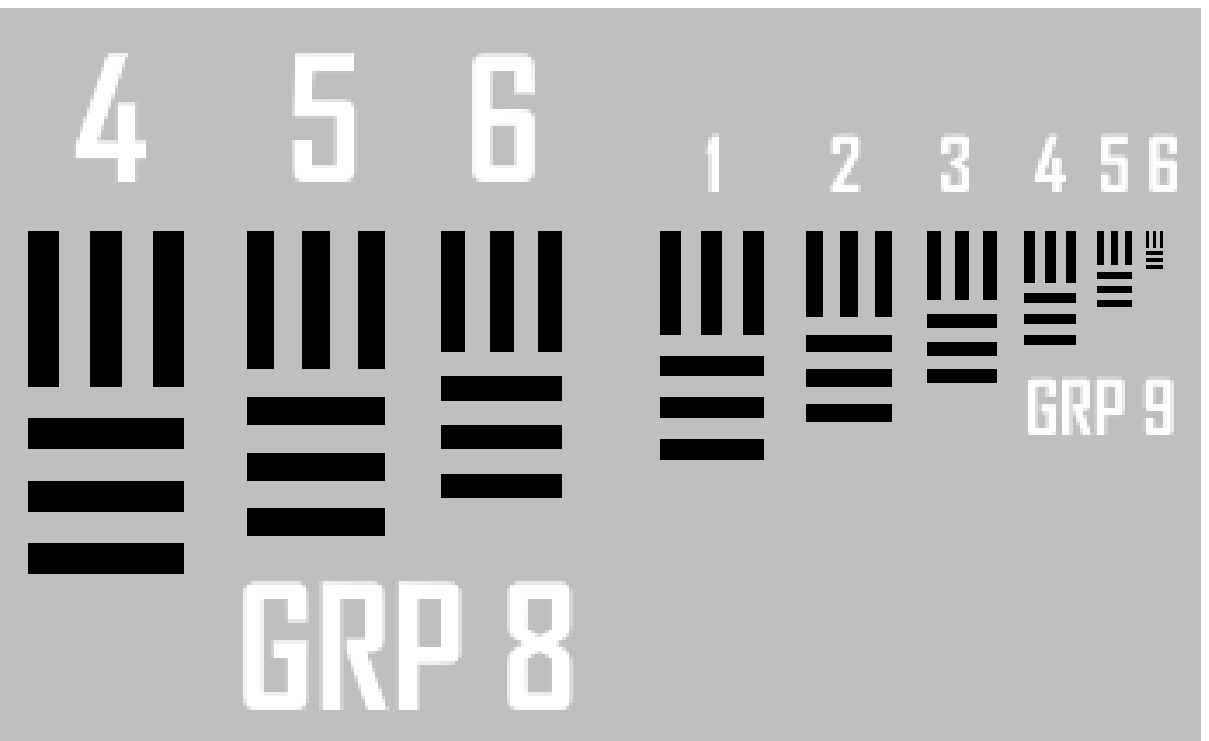}
    \linethickness{1pt}	
    \put(0,64){\color{black}{$\mathbf{a)}$}}
    \put(30,64){\color{black}{Original phase}}
    \put(76,7){\color{black}\line(1,0){16.5}}
    \put(77,10){\color{black}{\fontsize{6}{6}$\mathbf{50~\mu m}$}}
    \linethickness{0.5pt}	
    \put(53,41.5){\color{blue}\line(1,0){45}}
    \end{overpic}}
  \fbox{\begin{overpic}[height=3cm]{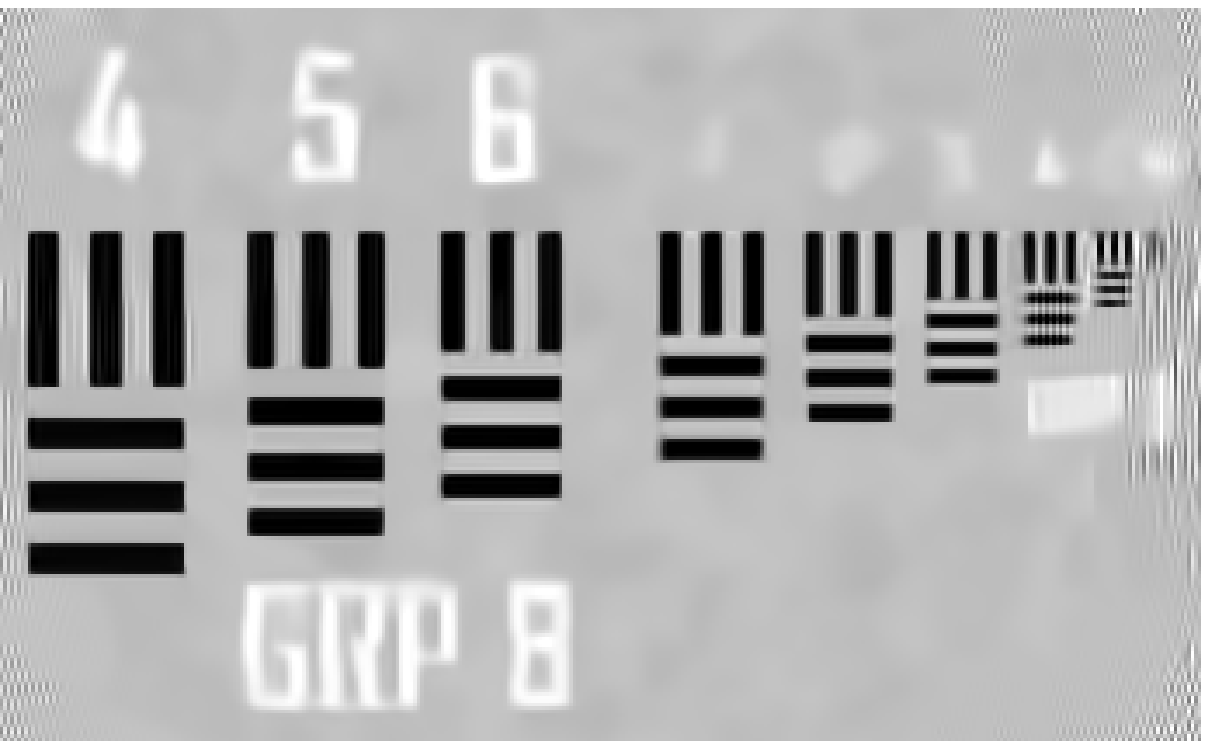}
    \linethickness{1pt}	
    \put(0,64){\color{black}{$\mathbf{b)}$}}
    \put(10,64){\color{black}{SR-SPAR (RRMSE=0.219)}}
    \put(76,7){\color{black}\line(1,0){16.5}}
    \put(77,10){\color{black}{\fontsize{6}{6}$\mathbf{50~\mu m}$}}
    \linethickness{0.5pt}	
    \put(53,41.5){\color{red}\line(1,0){45}}
    \end{overpic}}
  \fbox{\begin{overpic}[height=3cm]{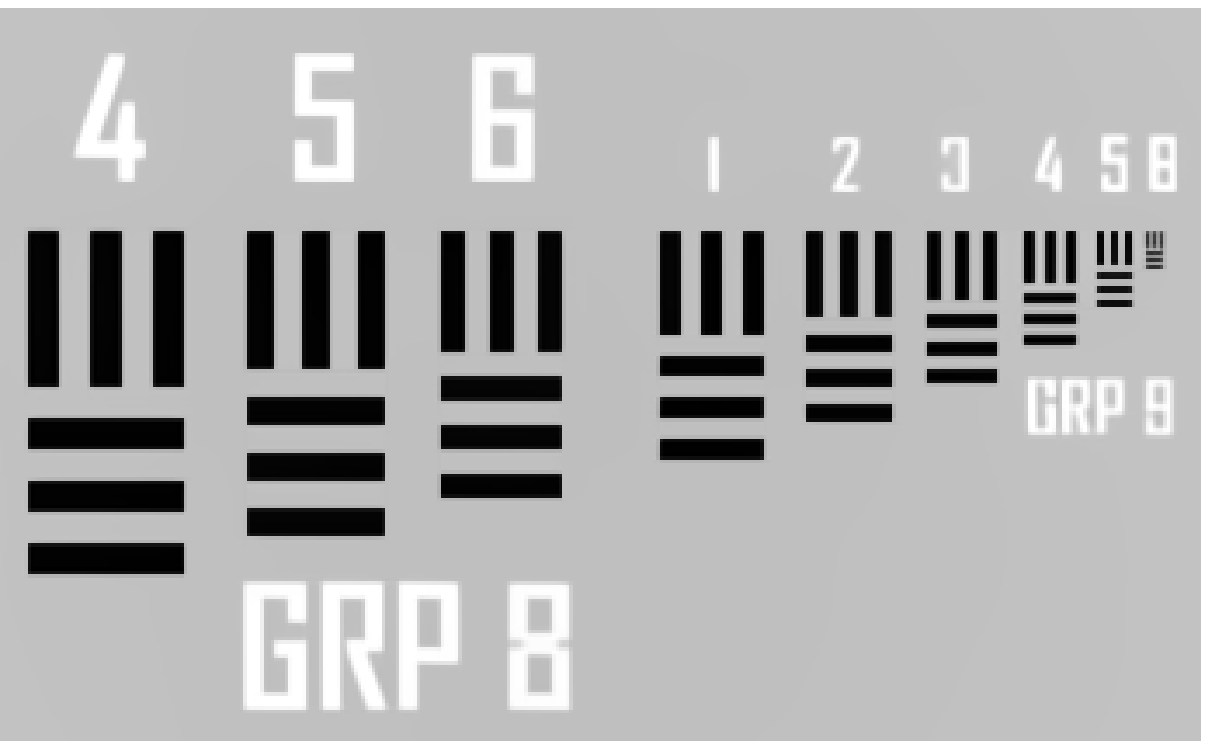}
    \linethickness{1pt}	
    \put(0,64){\color{black}{$\mathbf{c)}$}}
    \put(10,64){\color{black}{SSR-PR (RRMSE=0.073)}}
    \put(76,7){\color{black}\line(1,0){16.5}}
    \put(77,10){\color{black}{\fontsize{6}{6}$\mathbf{50~\mu m}$}}
    \linethickness{0.5pt}	
    \put(53,41.5){\color{green}\line(1,0){45}}
    \end{overpic}}
  \begin{overpic}[trim={33.7cm 3.9cm 13.5cm 1.1cm}, clip, height=3.1cm]{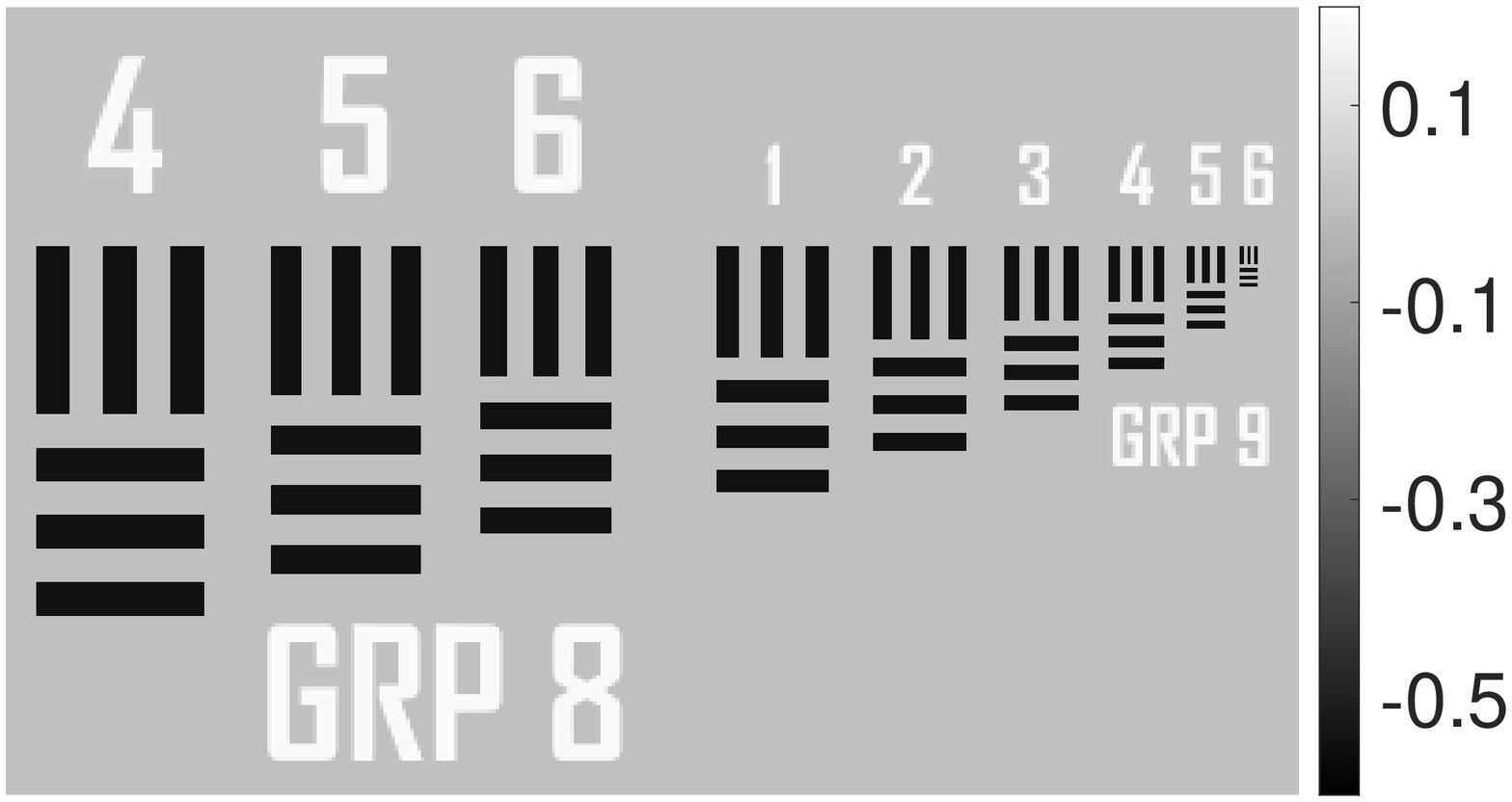}
    \put(29,24){\color{black}\fontsize{9}{9}\rotatebox{90}{$phase~[rad]$}}
    \end{overpic}
  \begin{overpic}[trim={6.5cm 6.8cm 1cm 0.5cm}, clip, width=\textwidth]{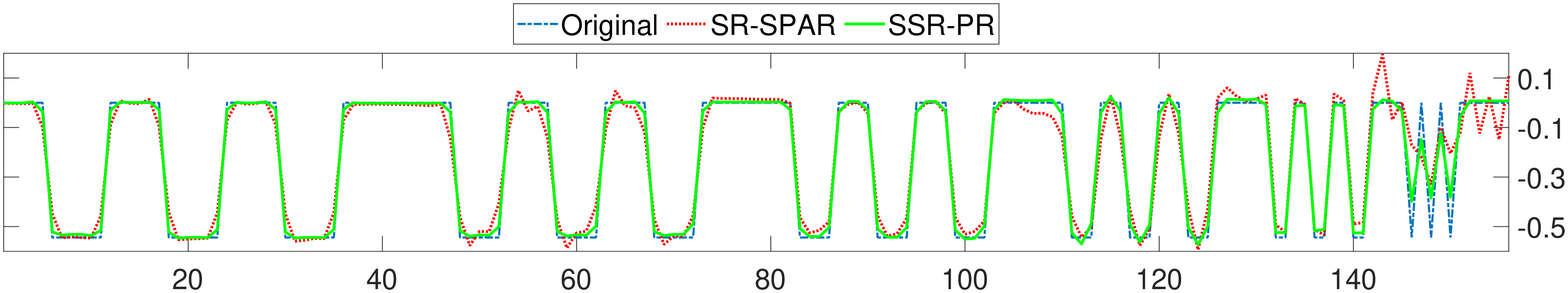}
    \put(96,6){\color{black}\fontsize{9}{9}\rotatebox{90}{$phase~[rad]$}}
    \put(43,-1){\color{black}\fontsize{9}{9}{$position~[px]$}}
  \end{overpic}
\caption{Phase reconstructions of the simulated phase-only USAF target, with their corresponding cross-sections. a) Original phase image, b) reconstructions with previous SR-SPAR\cite{kocsis2020single}, c) and the novel SSR-PR methods.}
\label{fig:results}
\end{figure}

To demonstrate the advantages of the SSR-PR approach and algorithm, we provide a set of simulations with USAF phase target as the object $u_{o}$ (Fig. \ref{fig:results}/a).
The object is assumed to follow the structure of a physical USAF phase target by simulating parallel etched lines for each elements per group.
The simulated etch depths of the lines are 100 nm, which are more than 5$\times$ smaller than the used illumination wavelength of $\lambda=532$ nm, corresponding to phase shift of only 
\begin{equation}
\label{eq:height2phase}
    \Delta\varphi = \frac{2 \pi (n-1) \Delta h}{\lambda}=\frac{2 \pi (1.4607-1) 100[nm]}{532[nm]} = 0.544~[rad].
    % \Delta \varhi = 2 \pi (n-1) \Delta h/\lambda=2 \pi (1.4607-1) 100[nm]/532[nm] = 0.544[rad]
\end{equation}
The wavefront modulation is made by a single stationary binary phase mask $M$. The mask parameter selection is based on our previous research\cite{kocsis2020single}. The experiments showed that the best performance is achieved if we select the mask pixels half the size of the sensor pixel's ($\Delta_{s}=3.5~\mu$m, $\Delta_{m}=1.75~\mu$m) and their maximum phase-shifts are taken close to $\pi$. Corresponding to these, the mask pixel heights are taken as binary random values with equal probabilities for 0 and 500 nm (0 and 2.72 rad). We also found that
For the intensity pattern generation it is desirable to take as small computational pixel size $\Delta_{c}$ as possible to give the best approximation of the continuous data. To simulate the behavior of the sensor, the patterns are downsampled by averaging every $r_{s}\times r_{s}$ sized patches and crop the required sensor size with $N_{1}\times N_{2}=2048\times2448$. However, a restriction occurs\cite{katkovnik2017computational} due to the discretization of the continuous wavefront, which limits the size of the computational matrices $N_{c}\times N_{c}$. The restriction is in connection with the summed propagation distance $d_{1}+d_{2}$, the computational pixel size $\Delta_{c}$, and the wavelength $\lambda$. The inequality defines the minimum size $N_{eff}$ of the computation matrices for effective sampling as
\begin{equation}
    r_{s}\cdot N_{c}\geq N_{eff}=(d_{1}+d_{2})\lambda/\Delta_{c}^2.
\end{equation}
The wavelength is given, and the distances are taken to imitate a physical scheme as $d_{1}=1$ mm and $d_{2}=8$ mm. Since the correct mask placing requires super-resolution factor power of two, and this factor is limited to $r_{s}=9.17$, the smallest computational pixel size can be taken as $\Delta_{c}=0.4375~\mu$m ($r_{s}=8$), resulting in $N_{eff}=25015$. To satisfy the inequality without peradventure, we took $r_{s}\cdot N_{x}=26000$. 
This way we have ideally modeled a high-resolution wavefront to be reconstructed with low-resolution observation using Lanczos-3 upsampling.

\subsection{Results}
\label{sec:results}

The phase reconstructions of the complex-valued object after 20 iterations are shown in Fig.~\ref{fig:results}. We have compared the SSR-PR reconstruction (c) with the original object (a) and the previous SR-SPAR method\cite{kocsis2020single} (b). Since the modulation mask causes corruption on the wavefront, the SR-SPAR requires a strong sparse filtering and narrow \textit{support constraint} window, which eliminates the small phase values.
Due to the improvements presented in Sec. \ref{sec:novel}, the SSR-PR approach provides significantly better reconstructions with 3$\times$ smaller RRMSE than the SR-SPAR, while preserving small phase values as well. The cross-sections of the phase values demonstrate the phase-correct super-resolution by resolving the 6th element of group 9. The line-thickness of this element is $0.875~\mu$m, which is 4$\times$ smaller than the sensor pixel size. Being able to resolve these small details is a significant advantage over the diffraction limited systems, since we can overcome the diffraction limited resolution $\Delta_{Abbe}$. This limit can be calculated by Abbe's criterion as
\begin{equation}
\label{eq:abbe}
    \Delta_{Abbe} = \frac{\lambda}{NA}\approx\frac{2d\lambda}{N\Delta_{s}}=\frac{2(d_{1}+d_{2})\lambda}{N\Delta_{s}}=1.34~[\mu m].
\end{equation}

\subsection{Error compensation demonstration}
\label{sec:errorcomp}

\begin{figure}[b!]
    \centering
    \begin{overpic}[trim={0cm 2.8cm 1cm 1cm}, clip, width=\textwidth]{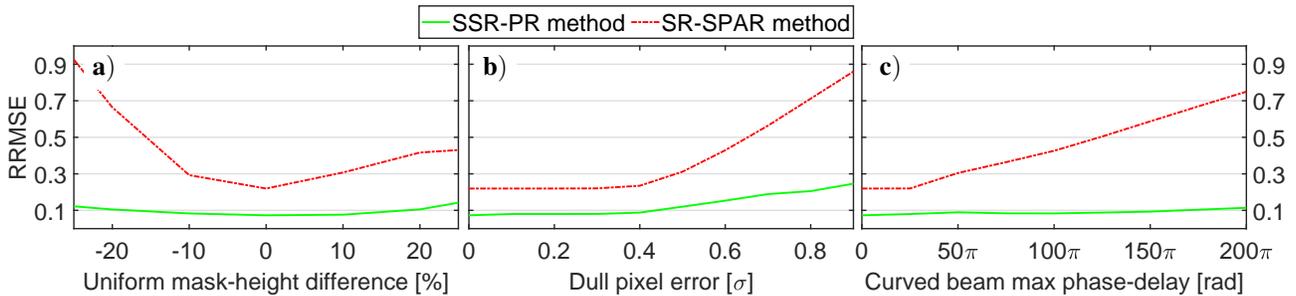}
    \put(7.5,17.5){\colorbox{white}{$\mathbf{a)}$}}
    \put(37,17.5){\colorbox{white}{$\mathbf{b)}$}}
    \put(67,17.5){\colorbox{white}{$\mathbf{c)}$}}
    \end{overpic}
    \caption{RRMSE of phase reconstructions with errors of the optical system. In a) and b) mask errors are assumed as uniform height-difference and dull pixels, while in c) the wavefront of the illumination beam is curved. }
    \label{fig:errors}
\end{figure}

One of the significant advantages of the SSR-PR is the robustness to setup errors between the ideal and real optical elements. 
Namely the modulation mask is assumed to be precisely known, although errors could appear due to the manufacturing tolerances. Furthermore, the theoretically plane illumination beam could have curvature on its phase. We simulated such errors separately and created the corresponding diffraction patterns. From these patterns, the object is retrieved and RRMSE of the previous and proposed reconstructions are shown in Fig.~\ref{fig:errors}. In the first case (Figure \ref{fig:errors}/a) we assumed a uniform height difference on all modulation mask pixels in the difference interval from -25\% to 25\%. In the second case (Fig. \ref{fig:errors}/b) dull pixel edges were simulated by using Gaussian filter with given $\sigma$ on the upsampled modulation mask. 
% In our simulations, $\Delta_{m} \times \Delta_{m} = 4 \Delta_{c} \times 4 \Delta_{c}$ with symmetric features even after the Gaussian filtering.
$\sigma<0.3$ means sharp edges, when all values correspond to the expected value. If $\sigma\geq0.3$ the values in a single mask pixel will change according to $\sigma$. The change in the value is a 20\% decrease at $\sigma=0.5$, and goes up to more than 67\% at $\sigma>1$. This means that at this point the single pixels will become dull bulges with less than 35\% of the expected value. 
In the third case (Figure \ref{fig:errors}/c) we simulate phase corresponding to the expanding wavefront, which is simulated as a approximation by adding spherical phase to the illumination beam $u_{b}$. The maximum phase shifts representing the wavefront curvature are shown in the x-axis. We assumed zero shifts at the edges and the maximum at the center. In Fig. \ref{fig:errors} we can see that the previous SR-SPAR method (red) is extremely sensitive to the errors and provides good results only when the system is near ideal. The SSR-PR method (green) is capable of compensating the appearing errors and significantly surpassing  SR-SPAR.

%% file: Sections/5_Physical_experiments.tex
\section{Physical experiments}

Three sets of physical experiments have been done. In the first one we aimed to demonstrate the super-resolution power of the proposed SSR-PR by reconstructing a calibrated USAF test-target. In the second set, the biomedical application possibility is demonstrated by resolving a static biological sample as Buccal Epithelial Cells without any special preparation.
In the third experiment, the video microscope application is demonstrated by capturing and reconstructing a video of a living dynamic biological sample as a moving single-celled eukaryote, so called protozoa. The reconstructions do not need any special preparation of the specimens.

In the case of static samples Digital Holographic Microscopy\cite{katkovnik2016high} (DHM) was used to verify the results. The DHM is using classical holographic principles and the off-axis hologram is recorded by a digital sensor.
% For DHM we used a laser illumination with wavelength of $\lambda=403$ nm, therefore the resulting phase differs from the one obtained with $\lambda=532$ nm. 
For proper comparison, the phases are converted into height values by rearranging Eq. \ref{eq:height2phase} as follows:
\begin{equation}
    \Delta h = \frac{\Delta \varphi \lambda}{2\pi(n-1)}.
\end{equation}
The refractive index $n$ of dry epithelial cells from oral cavity was taken from this paper\cite{belashov2017determination}.
% The mean refractive index of dry cells of oral-cavity epithelium was estimated as nc = 1.478
% Since the refractive index $n$ of the biological samples are unknown, we looked for close biomedical specimens and selected it corresponding to the average refractive index of red blood cells\cite{phillips2012measurement} as $n=1.4$.
% Since the refractive index $n$ of the biological samples are unknown, we assumed to be the same as the used glass plates (fused silica).

\begin{wrapfigure}{r}{0.55\textwidth}
\vspace{-5pt}
    \centering
    \begin{overpic}[trim={0cm 0cm 0cm 0cm}, clip, width=0.5\textwidth]{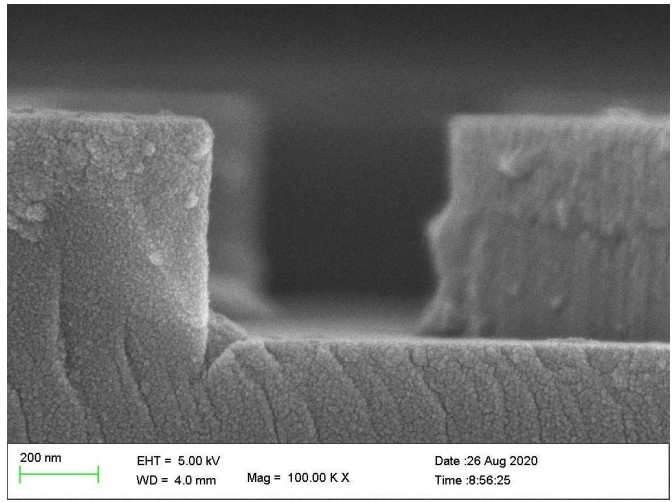}
    \end{overpic}
    \caption{Cut of the modulation mask.}
    \label{fig:maskpixel}
\end{wrapfigure}
For each experiment, a laser illumination was used with wavelength of $\lambda=532$ nm. The light modulation was made with a random binary phase mask, similar to the simulations. The previously used mask\cite{kocsis2020single} with 1 $\mu$m pixel size was replaced by a new mask with pixel size of $\Delta_{m}=1.73\mu$m. This size is half of the sensor's, therefore we can fit the mask better if the super-resolution factor is power of 2. As we stated before (Sec. \ref{sec:upsampling}) it is not worth taking $r_{s}>4$, since the calculation time will significantly increase, while the resolution will not change noticeably, so we selected it as $r_{s}=4$. The phase mask was manufactured with an electron beam lithography system on a fused silica glass. The phase-shift difference between the binary mask pixels are expected to be 2.72 rad (500 nm). however, by analyzing the physical mask by Scanning Electron Microscope (SEM), we found that the mask characterization includes the dullness (Sec. \ref{sec:errorcomp}) errors as shown in Fig. \ref{fig:maskpixel}. Furthermore, by measuring the surface of the mask, we found that a ca. 10\% uniform mask difference is present. 
The diffraction patterns were recorded with a CMOS sensor (FLIR Blackfly S BFS-U3-51S5M-BD) with the pixel size of $\Delta_{c}=3.45\mu$m, maximum resolution of $2448\times2048$, and a 12 bit dynamic range. The calculation was made on a computer with 32 GB of RAM and 3.41 GHz Intel(R) Core(TM) i7-6700 CPU. The software was written in MATLAB 2019b. For each set 10 iteration was made with field of view (FOV) of c.a. $1\times1$ $mm^{2}$. The algorithm was run on NVIDIA GeForce GTX 4050 Ti GPU and took around 50 seconds per iteration.
% The propagation was run on NVIDIA GeForce GTX 4050 Ti GPU and took around 4.5 seconds per iteration. Some parts of the BM3D was implemented as mexw64 files, so they run only on CPU and took around 74 seconds per iteration. This time could be lowered by lowering the size of the filtered area, but it will result in smaller FOV.

\subsection{USAF target}

\definecolor{ao(english)}{rgb}{0.0, 0.5, 0.0}
\definecolor{cyan}{rgb}{0.0, 1.0, 1.0}

\begin{figure}[t!]
    \centering
    \begin{overpic}[trim={2.3cm 1.6cm 2.3cm .8cm}, clip, height=3cm]{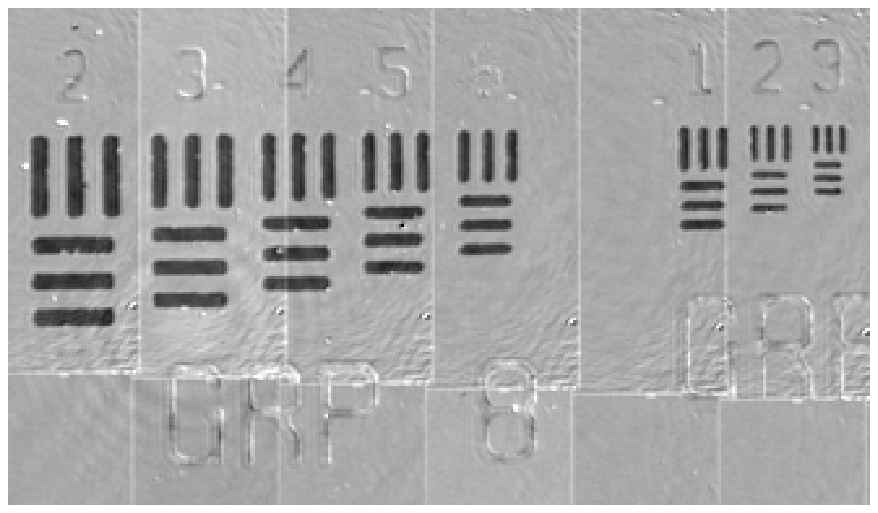}
    \put(0,60){\color{black}{$\mathbf{a)}$}}
    \put(35,60){\color{black}{DH method}}
    \linethickness{1pt}	
    \put(77,5){\color{black}\line(1,0){17.5}}
    \put(78,7){\color{black}{\fontsize{8}{8}$\mathbf{50~\mu m}$}}
    \linethickness{0.5pt}	
    \put(1,43){\color{red}\line(1,0){97}}
    \end{overpic}
    \begin{overpic}[trim={2.3cm 1.6cm 2.3cm .8cm}, clip, height=3cm]{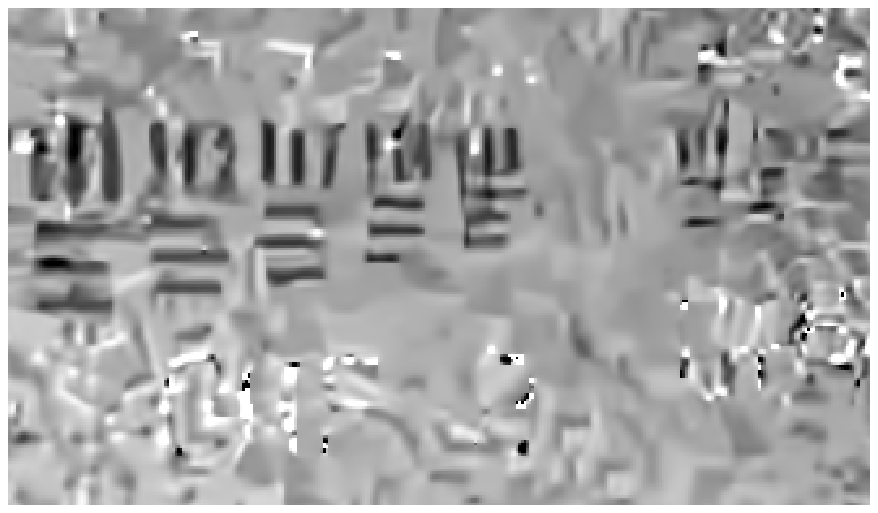}
    \put(0,60){\color{black}{$\mathbf{b)}$}}
    \put(25,60){\color{black}{SR-SPAR method}}
    \linethickness{1pt}	
    \put(77,5){\color{black}\line(1,0){17.5}}
    \put(78,7){\color{black}{\fontsize{8}{8}$\mathbf{50~\mu m}$}}
    \linethickness{0.5pt}	
    \put(1,43){\color{cyan}\line(1,0){97}}
    \end{overpic}
    \begin{overpic}[trim={2.3cm 1.6cm 2.3cm .8cm}, clip, height=3cm]{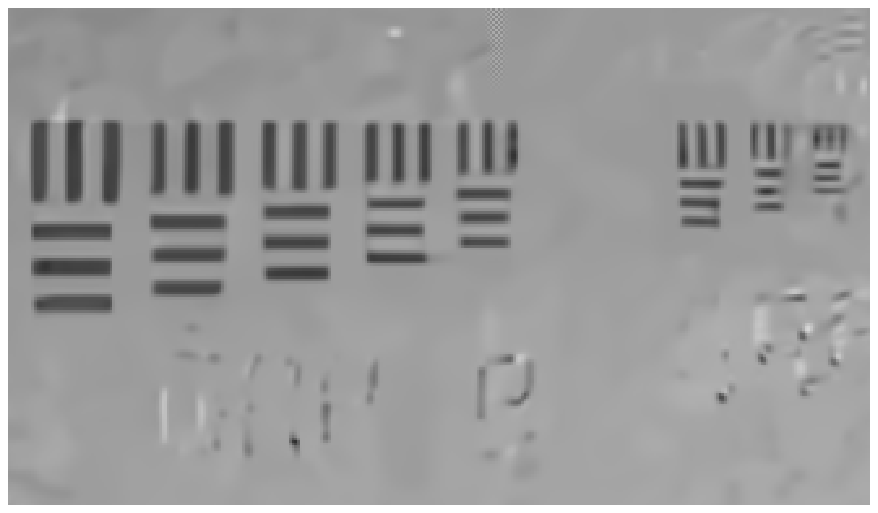}
    \put(0,60){\color{black}{$\mathbf{c)}$}}
    \put(25,60){\color{black}{SSR-PR method}}
    \linethickness{1pt}	
    \put(77,5){\color{black}\line(1,0){17.5}}
    \put(78,7){\color{black}{\fontsize{8}{8}$\mathbf{50~\mu m}$}}
    \linethickness{0.5pt}	
    \put(1,43){\color{green}\line(1,0){97}}
    \end{overpic}
    \begin{overpic}[trim={34cm 3.9cm 11.5cm 1.1cm}, clip, height=3.1cm]{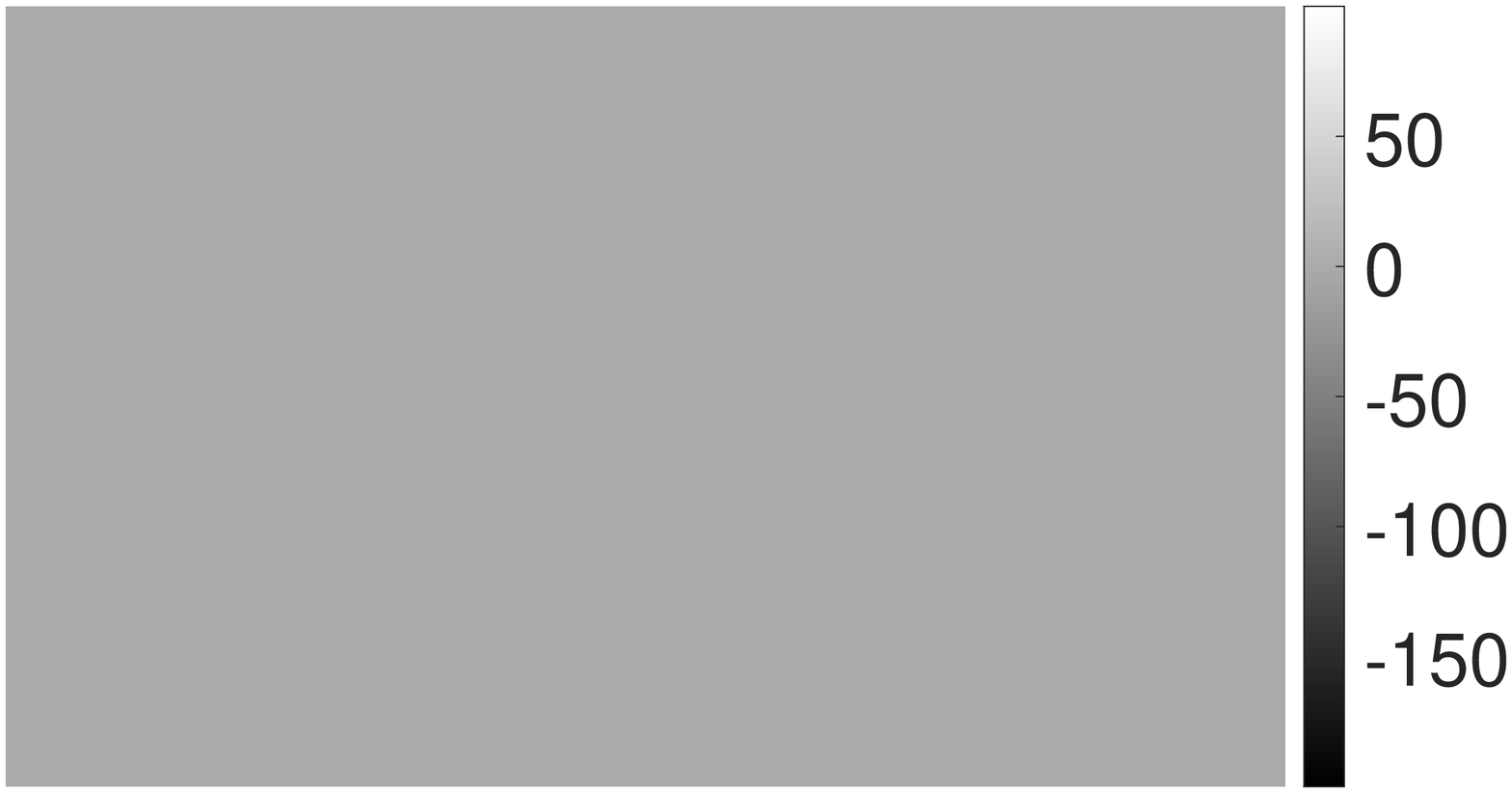}
    \put(29,24){\color{black}\fontsize{9}{9}\rotatebox{90}{$height~[nm]$}}
    \end{overpic}
    \linebreak
    \begin{overpic}[trim={6.5cm 6.5cm 1.5cm .5cm}, clip, width=\textwidth]{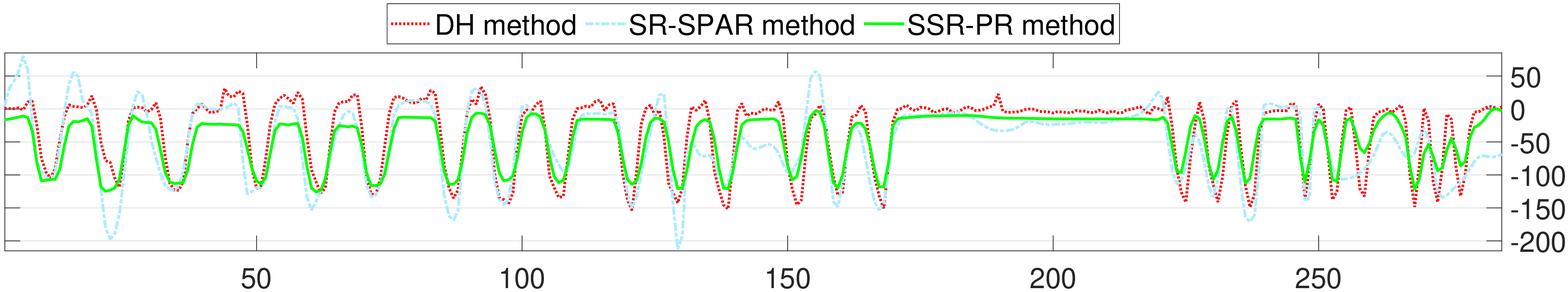}
    \put(97,6){\color{black}\fontsize{9}{9}\rotatebox{90}{$height~[nm]$}}
    \put(43,-1){\color{black}\fontsize{9}{9}{$position~[\mu m]$}}
  \end{overpic}
  \caption{Height maps of the calibrated USAF Phasefocus target\cite{godden2016phase} from the phase reconstructions with their corresponding cross-sections. a) Digital holographic method, b) previous SR-SPAR method, c) SSR-PR method.}
    \label{fig:USAFreco}
\end{figure}

The reconstructions of the resolution target (Phasefocus PFPT01-16-127\cite{godden2016phase}) are shown in Figure \ref{fig:USAFreco}, where the etch-depth for each line is 127 nm. We have compared the result of DHM (Fig. \ref{fig:USAFreco}/a), the reconstruction using the previous SR-SPAR (Fig. \ref{fig:USAFreco}/b) and the novel SSR-PR methods (Fig. \ref{fig:USAFreco}/c).
The SR-SPAR provides correction in the phase compared to the single-exposure maskless case, however, as shown in the simulations (Sec. \ref{sec:errorcomp}), it is very sensitive to the errors of the optical elements. These error cause corruption in the reconstructions, which can be eliminated only by an effective compensation. We can see that the compensation based SSR-PR approach is capable to resolve even the smallest group of the object with line-width of 2 $\mu$m, which is almost 2$\times$ smaller than the sensor pixel size. Using $CPO$ results in a much clearer reconstruction of the object, while providing significantly better phase-values than the traditional maskless phase retrieval. 
%With more adjustment the previous method is also capable of super-resolution imaging \cite{kocsis2020single}, but the results will be very noisy, which limits the resolving power.

\subsection{Static biological sample}

\begin{figure}[b!]
    \centering
    \begin{overpic}[trim={2.3cm 1.6cm 2.3cm .8cm}, clip, height=3.3cm]{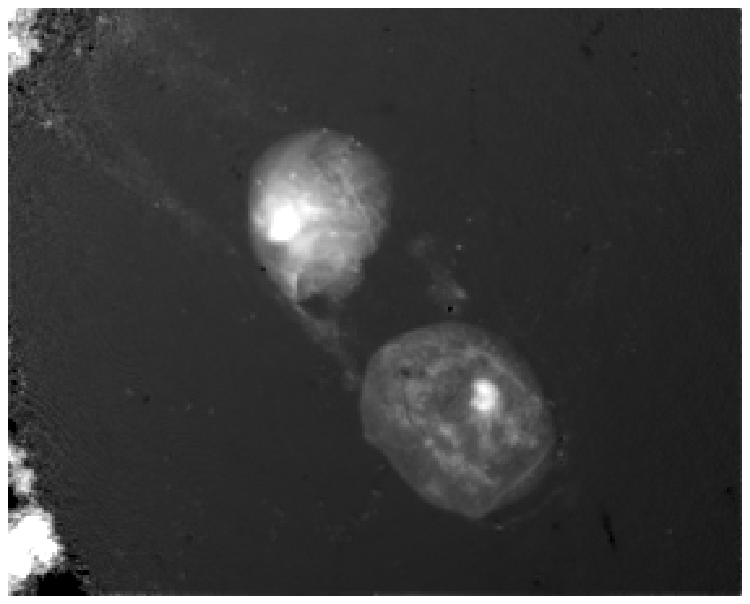}
    \put(0,83){\color{black}{$\mathbf{a)}$}}
    \put(30,83){\color{black}{DH method}}
    \linethickness{1pt}	
    \put(75,7){\color{white}\line(1,0){20.5}}
    \put(75,9){\color{white}{\fontsize{8}{8}$\mathbf{50~\mu m}$}}
    \linethickness{0.5pt}	
    \put(43,26.5){\color{red}\line(1,0){35}}
    \end{overpic}
    \begin{overpic}[trim={2.3cm 1.6cm 2.3cm .8cm}, clip, height=3.3cm]{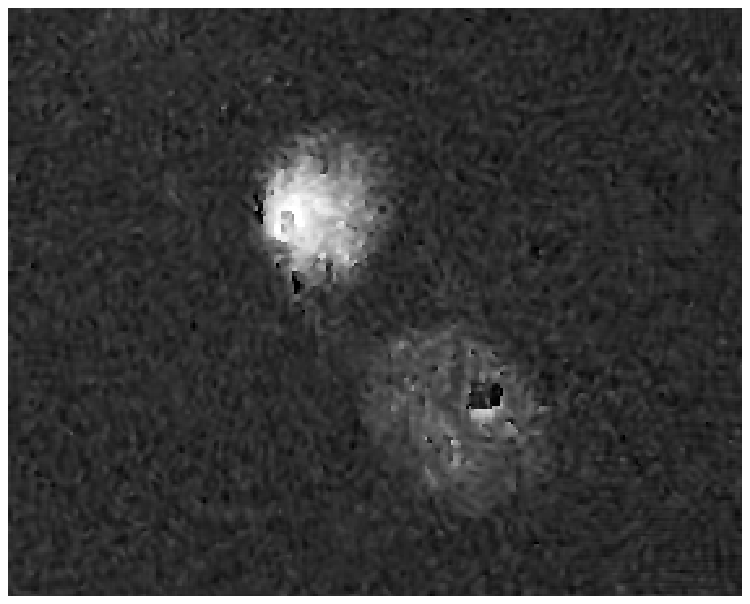}
    \put(0,83){\color{black}{$\mathbf{b)}$}}
    \put(20,83){\color{black}{SR-SPAR method}}
    \linethickness{1pt}	
    \put(75,7){\color{white}\line(1,0){20.5}}
    \put(75,9){\color{white}{\fontsize{8}{8}$\mathbf{50~\mu m}$}}
    \linethickness{0.5pt}	
    \put(43,26.5){\color{cyan}\line(1,0){35}}
    \end{overpic}
    \begin{overpic}[trim={2.3cm 1.6cm 2.3cm .8cm}, clip, height=3.3cm]{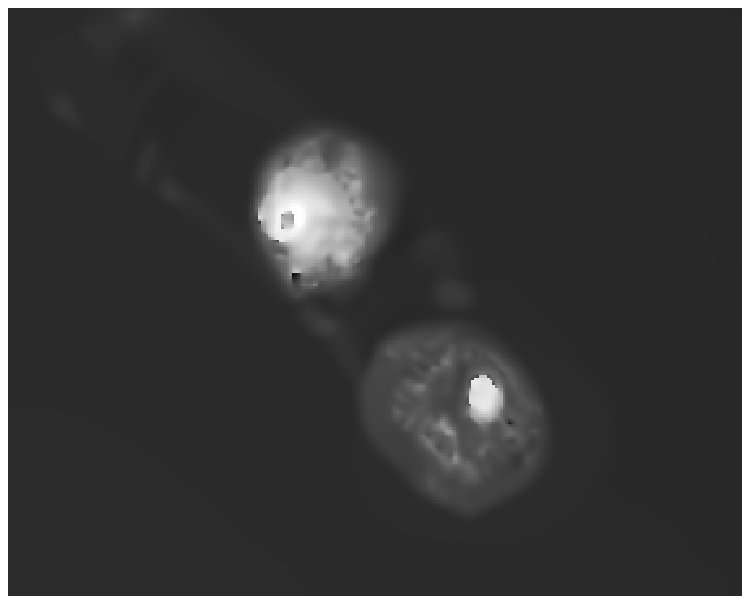}
    \put(0,83){\color{black}{$\mathbf{c)}$}}
    \put(20,83){\color{black}{SSR-PR method}}
    \linethickness{1pt}	
    \put(75,7){\color{white}\line(1,0){20.5}}
    \put(75,9){\color{white}{\fontsize{8}{8}$\mathbf{50~\mu m}$}}
    \linethickness{0.5pt}	
    \put(43,26.5){\color{green}\line(1,0){35}}
    \end{overpic}
    \begin{overpic}[trim={33.6cm 2.8cm 11.5cm 1.4cm}, clip, height=3.3cm]{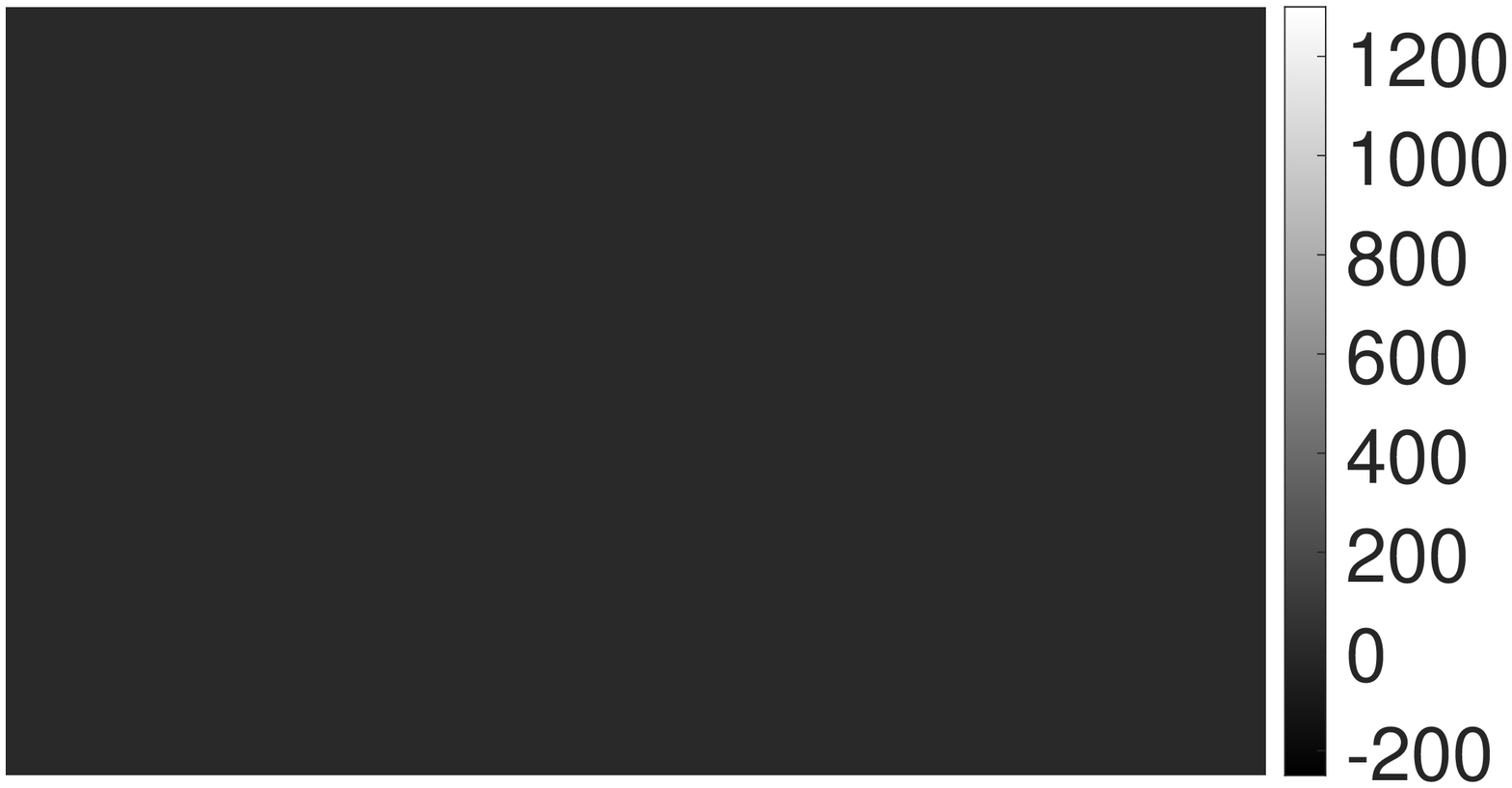}
    \put(32,24){\color{black}\fontsize{9}{9}\rotatebox{90}{$height~[nm]$}}
    \end{overpic}
    \space \space \space
    \begin{overpic}[trim={6.5cm 6.5cm 1.5cm .5cm}, clip, width=\textwidth]{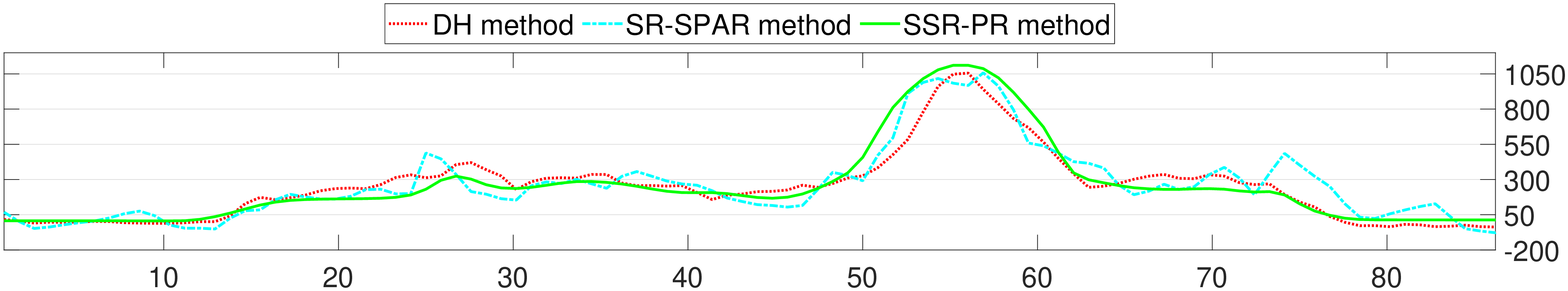}
    \put(97,6){\color{black}\fontsize{9}{9}\rotatebox{90}{$height~[nm]$}}
    \put(43,-1){\color{black}\fontsize{9}{9}{$position~[\mu m]$}}
  \end{overpic}
    \caption{Height maps of Buccal Epithelial Cells from the unwrapped phase reconstructions with their corresponding cross-sections. a) Digital holographic method, b) previous SR-SPAR method, c) SSR-PR method.}
    \label{fig:CELLreco}
\end{figure}

The reconstructions of the biological sample (buccal epithelial cells) are shown in Fig. \ref{fig:CELLreco}. The sample was taken from the mouth and applied on a sample slide without any preceding preparation. Similar to the previous experiment with the USAF target, we have compared the result of DHM (Fig. \ref{fig:CELLreco}/a), the reconstruction using the previous SR-SPAR (Fig. \ref{fig:CELLreco}/b) and the novel SSR-PR methods (Fig. \ref{fig:CELLreco}/c). The phase reconstructions were wrapped, therefore Phase Unwrapping via MAx flows\cite{valadao2007puma} (PUMA) algorithm was used to unwrap them. We can see that the SR-SPAR already provides phase-correct reconstruction, however, the result is generally noisy. It follows that the quality is inadequate and the unwrapping is challenging. 
In SSR-PR, the noises are compensated and filtered out properly, which results not only better quality but correct unwrapping as well. Comparing the reconstructions, we can see that the SSR-PR offers quality phase imaging with a good correspondence to the DH system - with a much simpler optical setup.

\subsection{Dynamic biological sample}

\begin{figure}[t!]
    \centering
    \begin{overpic}[height=3.1cm]{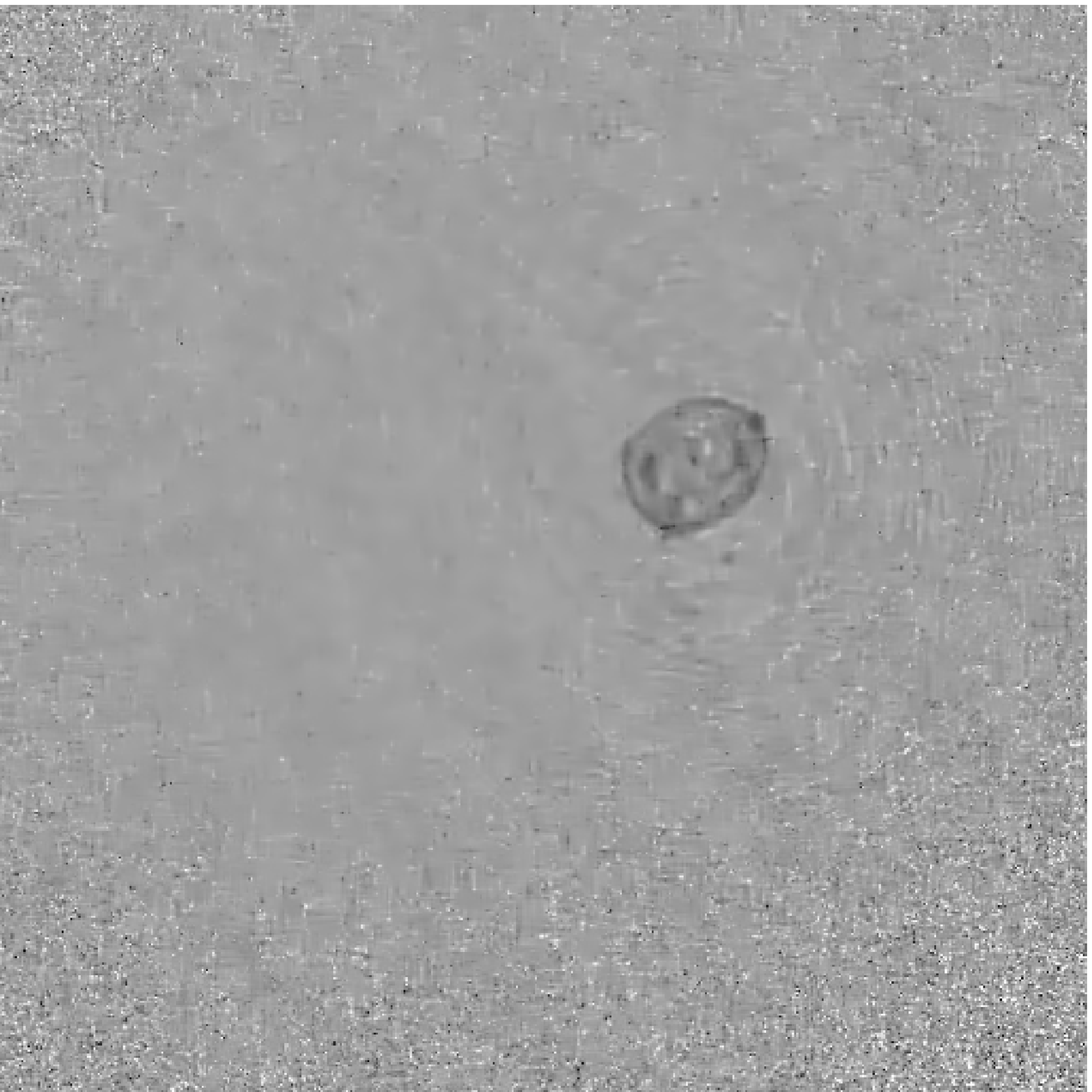}
    \linethickness{1pt}	
    \put(68,10){\color{black}\line(1,0){22.5}}
    \put(63,12){\color{black}{\fontsize{8}{8}$\mathbf{100~\mu m}$}}
    \end{overpic}
    \begin{overpic}[height=3.1cm]{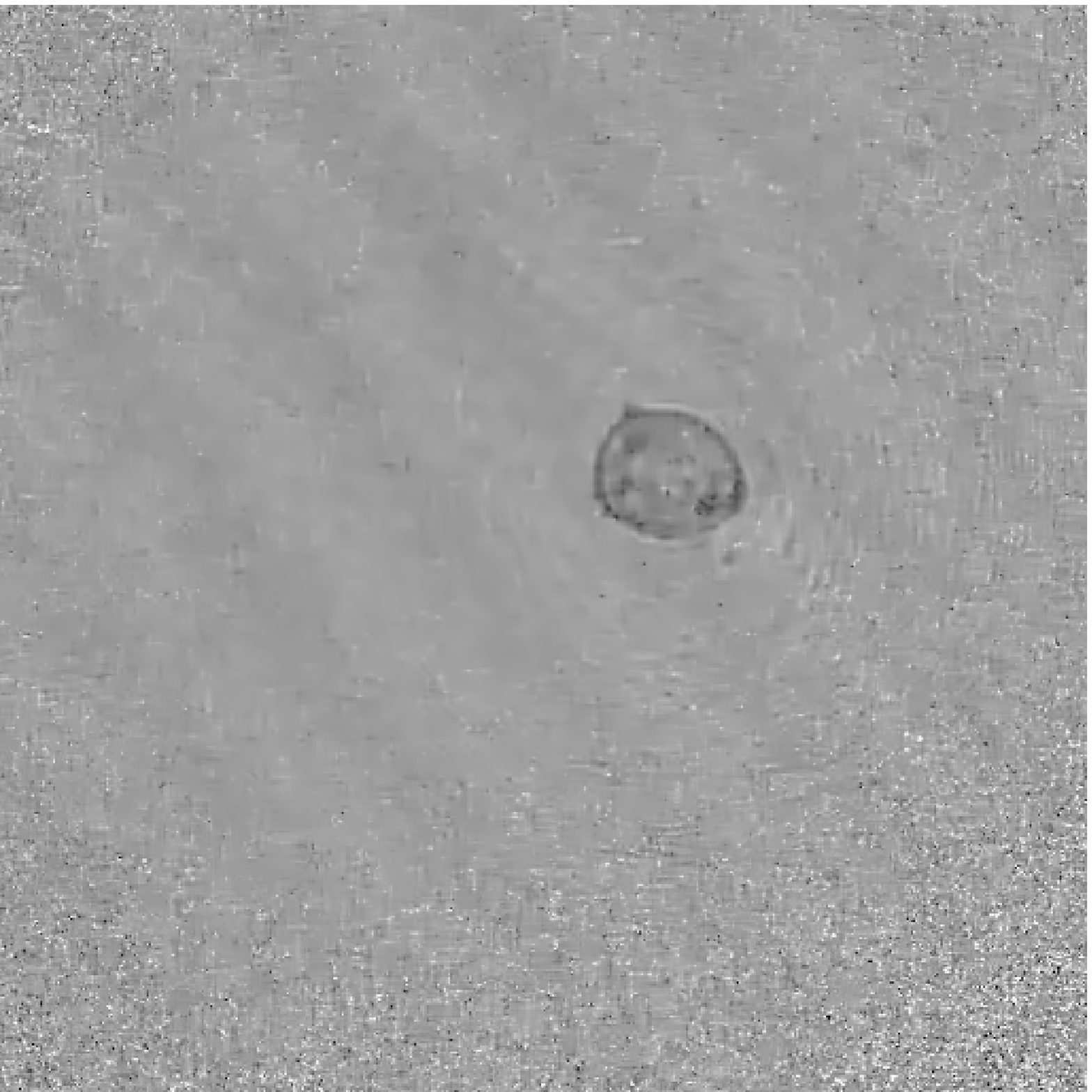}
    \linethickness{1pt}	
    \put(68,10){\color{black}\line(1,0){22.5}}
    \put(63,12){\color{black}{\fontsize{8}{8}$\mathbf{100~\mu m}$}}
    \end{overpic}
    \begin{overpic}[height=3.1cm]{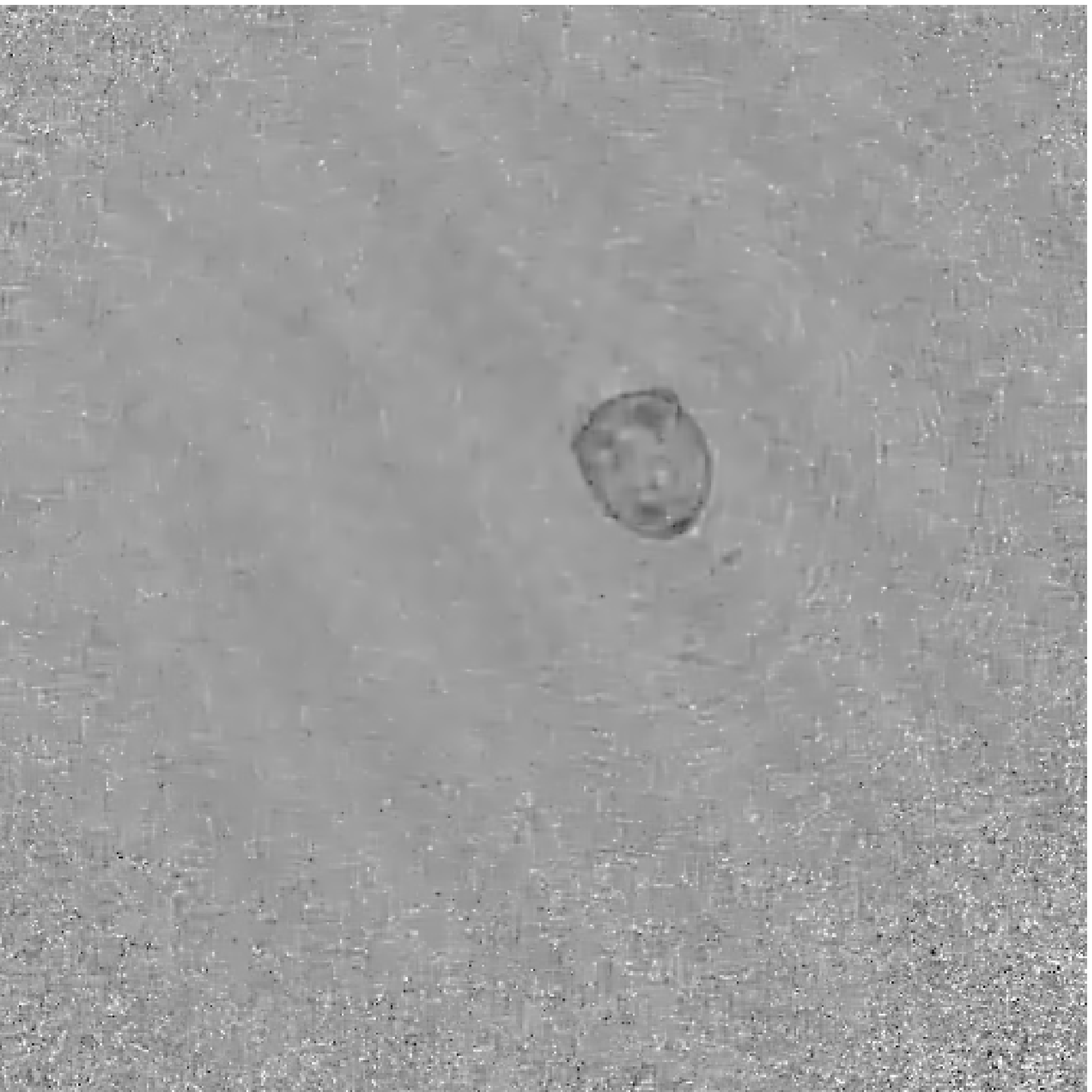}
    \linethickness{1pt}	
    \put(68,10){\color{black}\line(1,0){22.5}}
    \put(63,12){\color{black}{\fontsize{8}{8}$\mathbf{100~\mu m}$}}
    \end{overpic}
    \begin{overpic}[height=3.1cm]{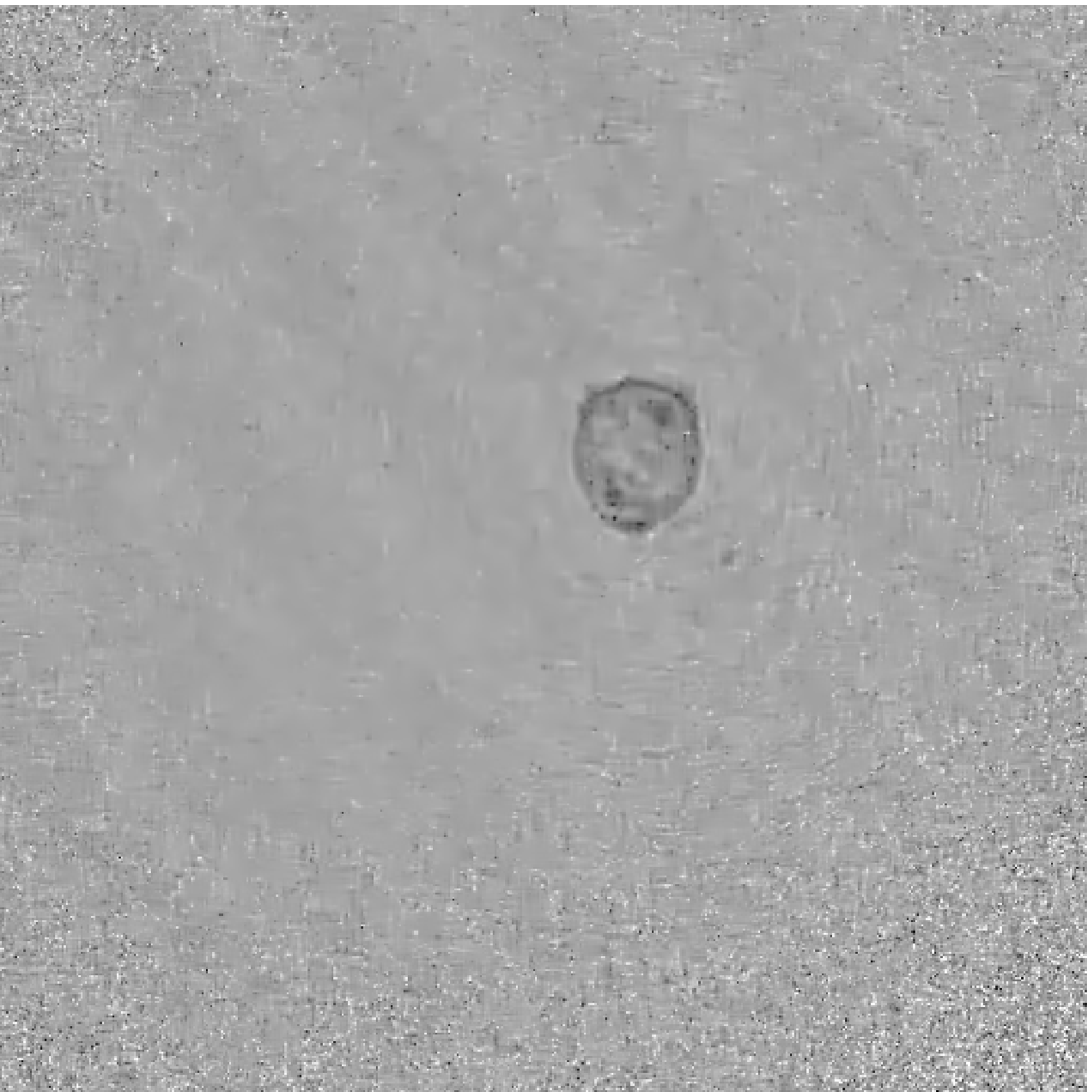}
    \linethickness{1pt}	
    \put(68,10){\color{black}\line(1,0){22.5}}
    \put(63,12){\color{black}{\fontsize{8}{8}$\mathbf{100~\mu m}$}}
    \end{overpic}
    \begin{overpic}[height=3.1cm]{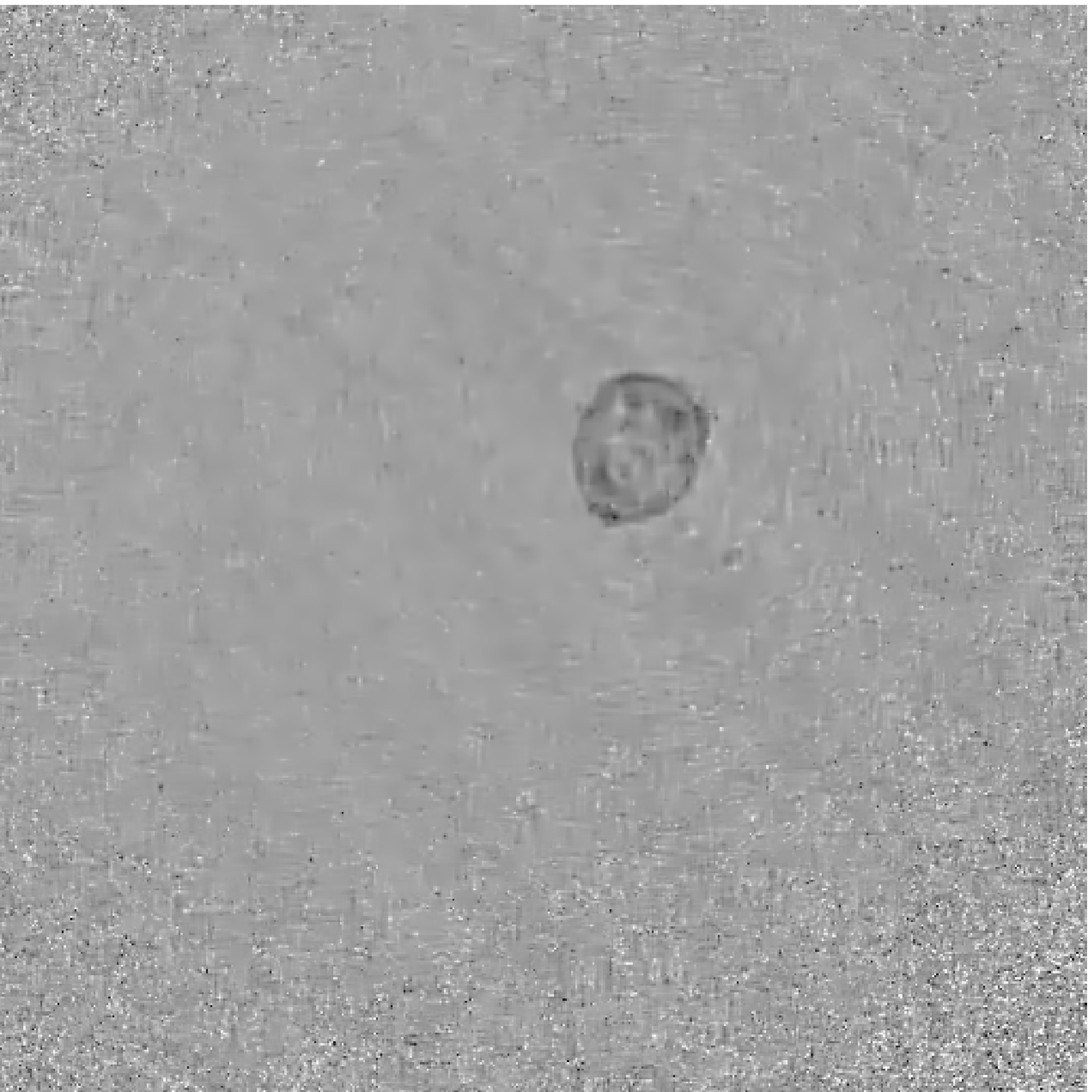}
    \linethickness{1pt}	
    \put(68,10){\color{black}\line(1,0){22.5}}
    \put(63,12){\color{black}{\fontsize{8}{8}$\mathbf{100~\mu m}$}}
    \end{overpic}
    \begin{overpic}[trim={33.6cm 3.2cm 11.5cm 1.4cm}, clip, height=3.1cm]{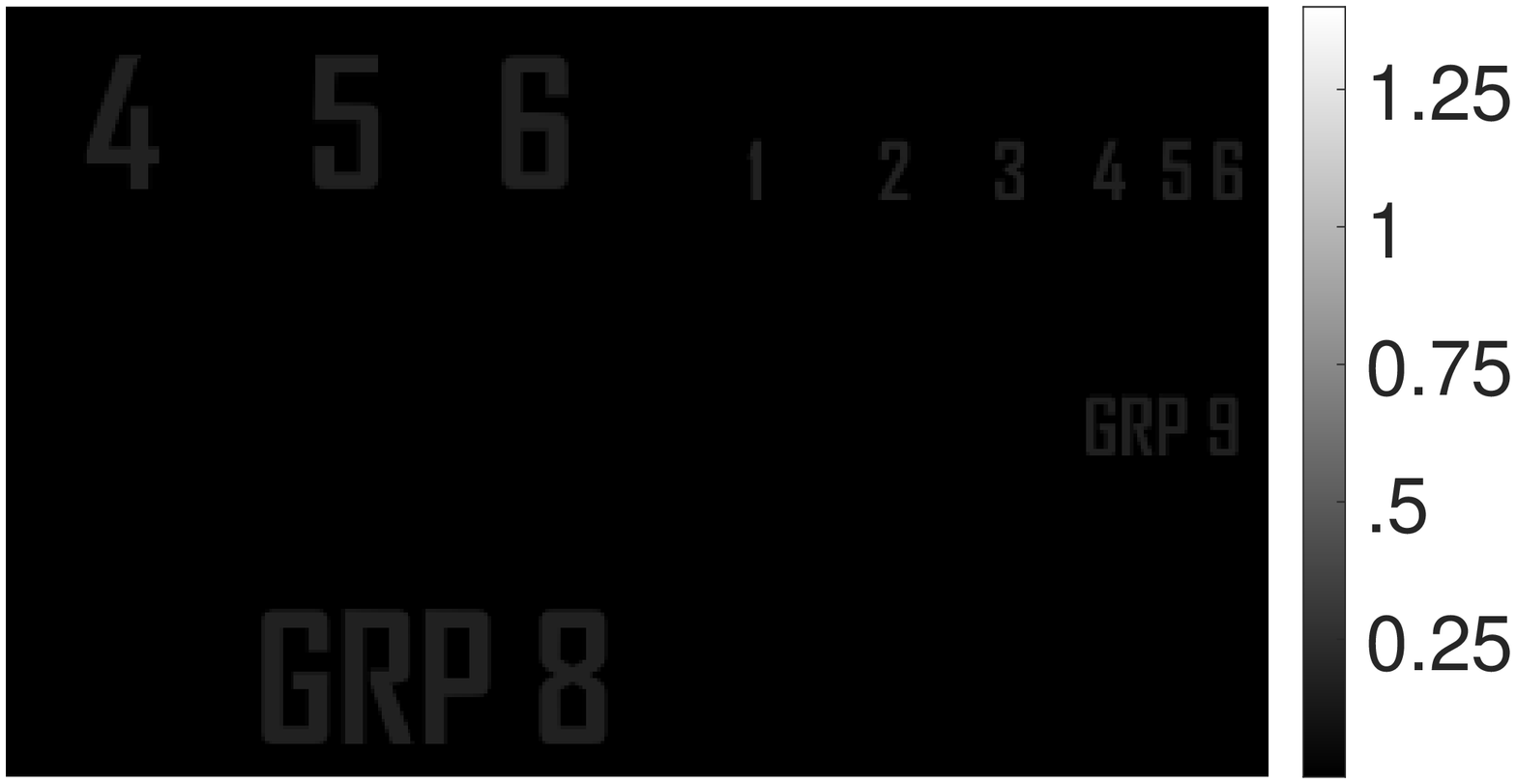}
    \put(32,24){\color{black}\fontsize{9}{9}\rotatebox{90}{$amplitude$}}
    \end{overpic}
    \begin{overpic}[height=3.1cm]{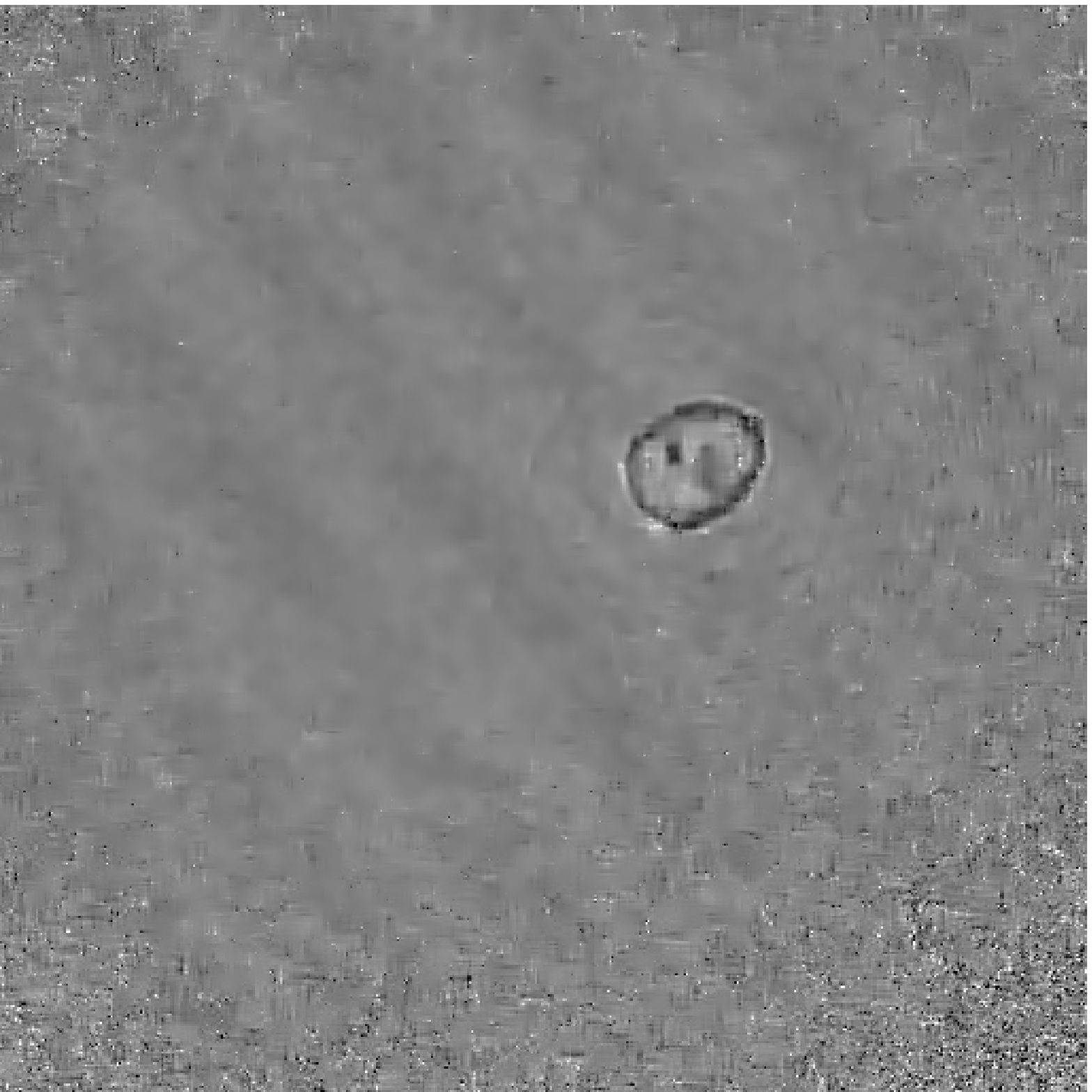}
    \linethickness{1pt}	
    \put(68,10){\color{black}\line(1,0){22.5}}
    \put(63,12){\color{black}{\fontsize{8}{8}$\mathbf{100~\mu m}$}}
    \end{overpic}
    \begin{overpic}[height=3.1cm]{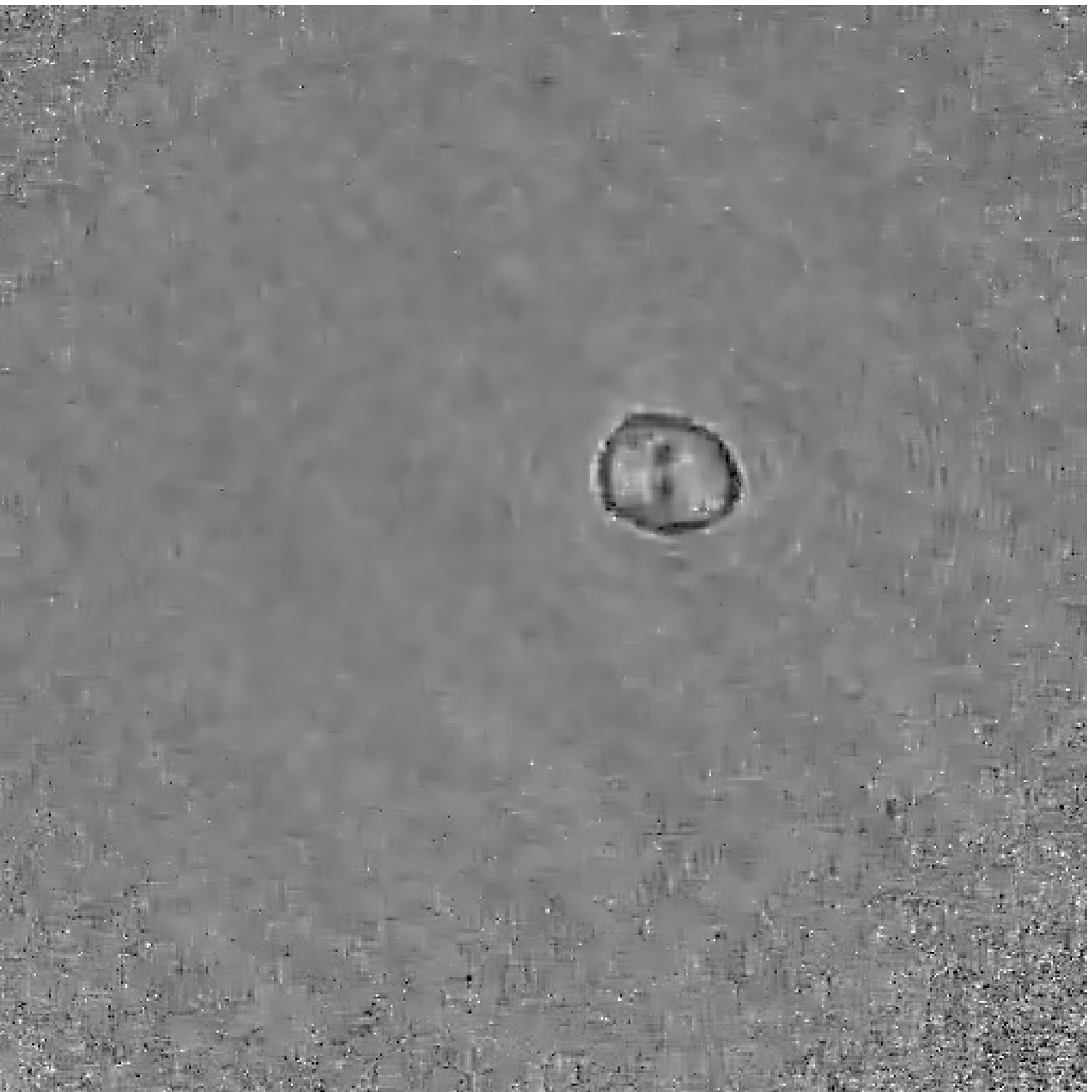}
    \linethickness{1pt}	
    \put(68,10){\color{black}\line(1,0){22.5}}
    \put(63,12){\color{black}{\fontsize{8}{8}$\mathbf{100~\mu m}$}}
    \end{overpic}
    \begin{overpic}[height=3.1cm]{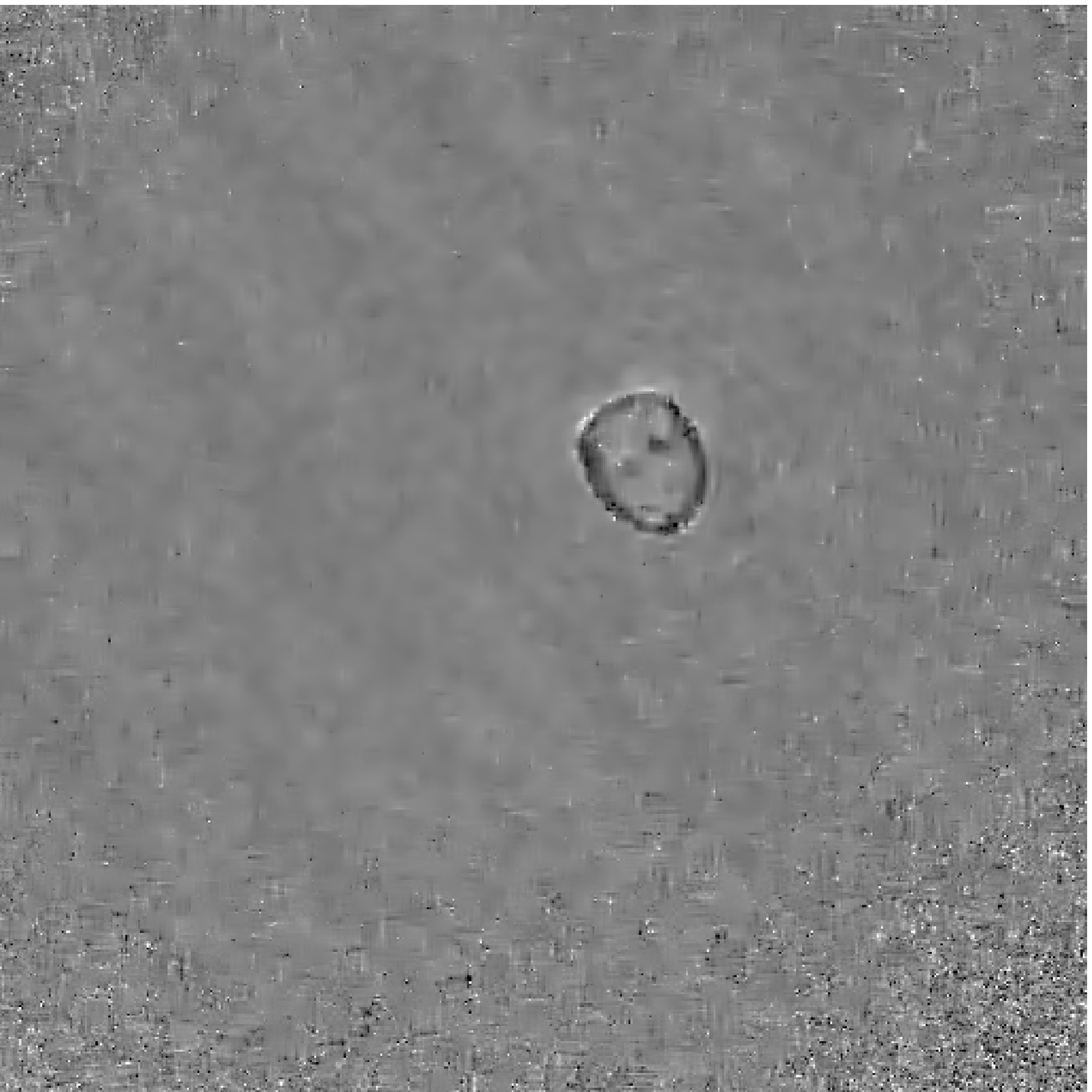}
    \linethickness{1pt}	
    \put(68,10){\color{black}\line(1,0){22.5}}
    \put(63,12){\color{black}{\fontsize{8}{8}$\mathbf{100~\mu m}$}}
    \end{overpic}
    \begin{overpic}[height=3.1cm]{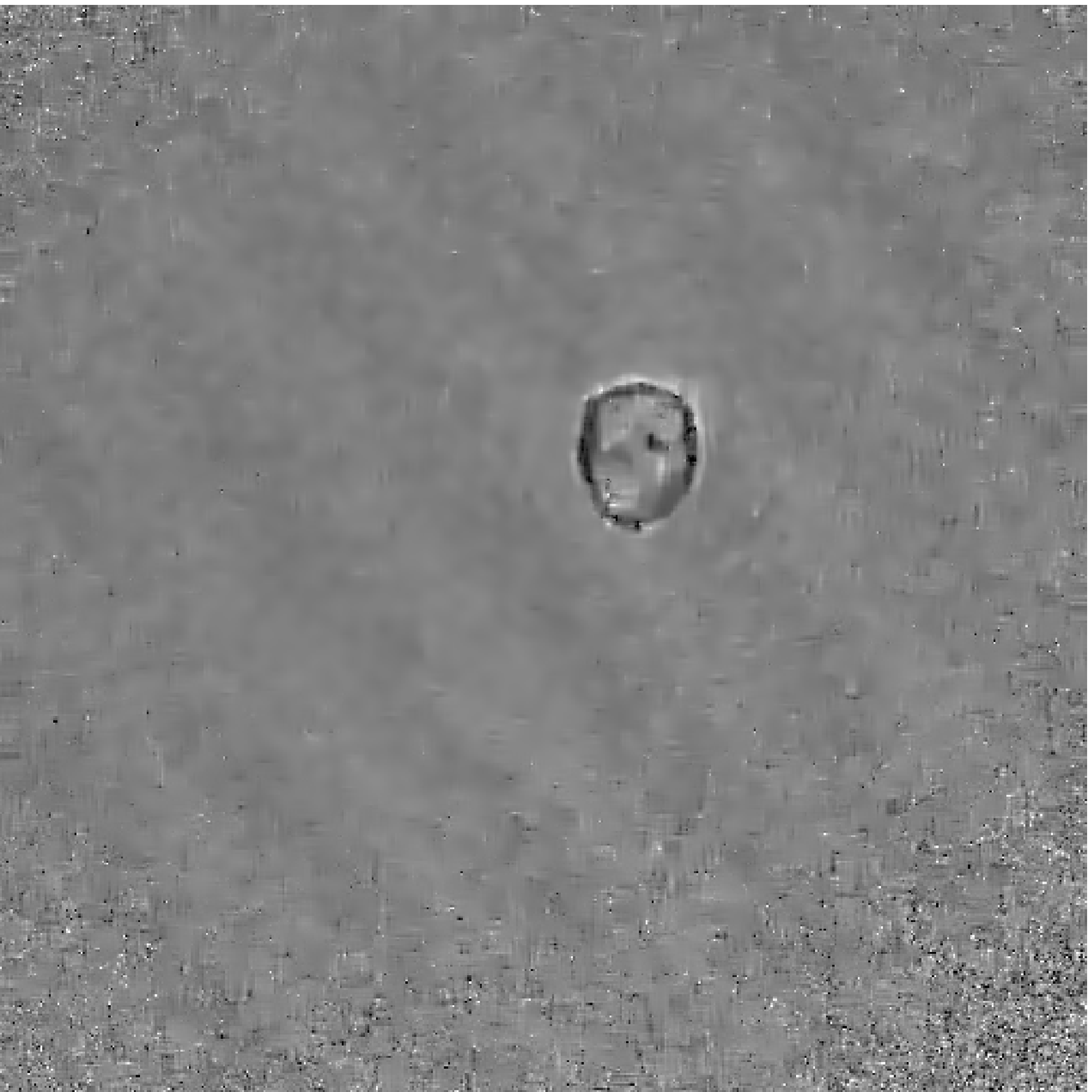}
    \linethickness{1pt}	
    \put(68,10){\color{black}\line(1,0){22.5}}
    \put(63,12){\color{black}{\fontsize{8}{8}$\mathbf{100~\mu m}$}}
    \end{overpic}
    \begin{overpic}[height=3.1cm]{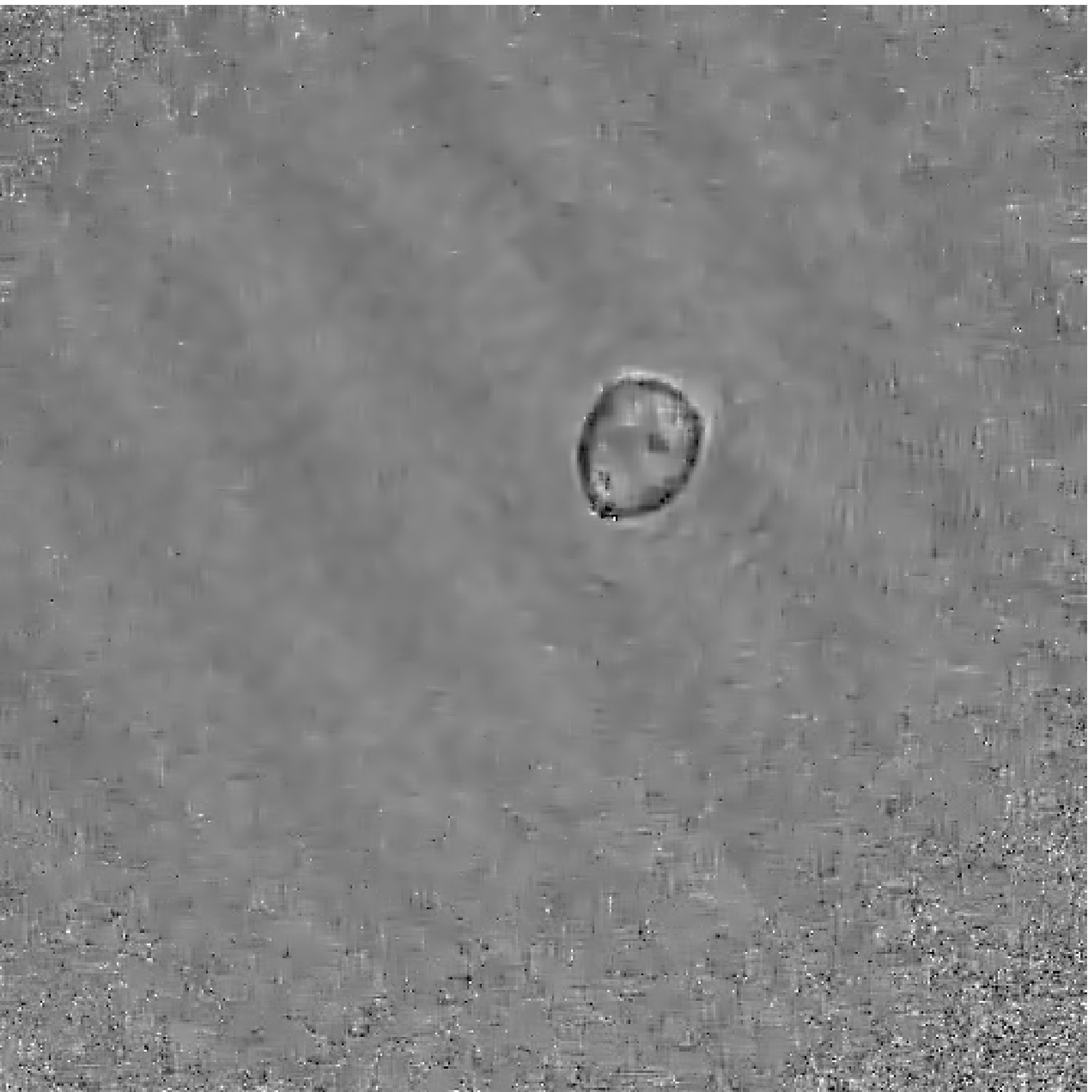}
    \linethickness{1pt}	
    \put(68,10){\color{black}\line(1,0){22.5}}
    \put(63,12){\color{black}{\fontsize{8}{8}$\mathbf{100~\mu m}$}}
    \end{overpic}
    \begin{overpic}[trim={33.6cm 3.2cm 11.5cm 1.4cm}, clip, height=3.1cm]{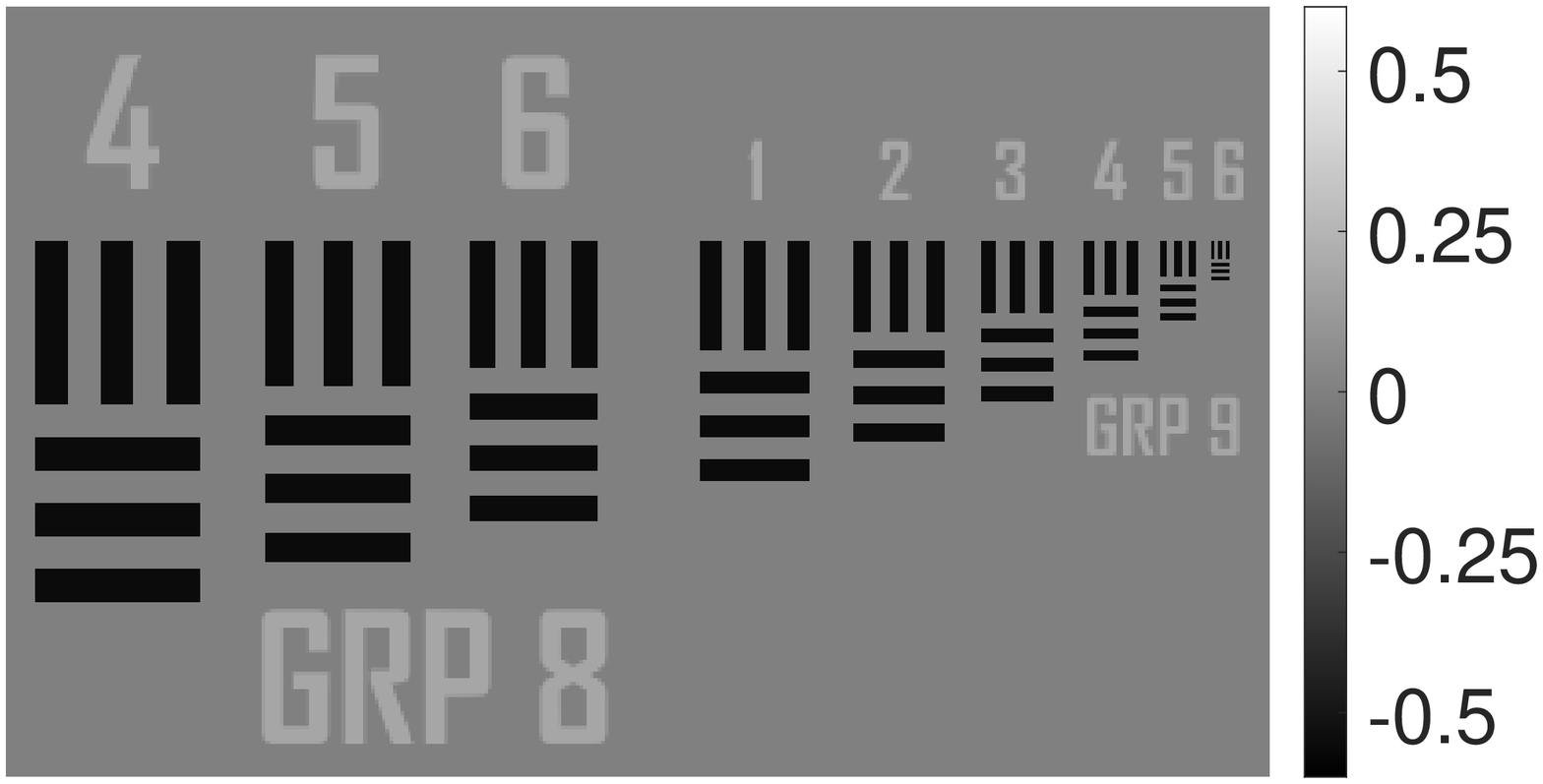}
    \put(34,24){\color{black}\fontsize{9}{9}\rotatebox{90}{$phase~[rad]$}}
    \end{overpic}
    \caption{Frames from complex-valued dynamic object video reconstruction. Amplitude (top row) and phase (bottom row) reconstructions of a moving single-celled eukaryote. The \href{https://tuni-my.sharepoint.com/:v:/g/personal/peter_kocsis_tuni_fi/EZBhbTmu9zZJsSIIsmPLOk0BjVDk5I4YcN3V_sf0BAS6-Q?e=JTvZEW}{video footage} consists 287 frames through 10 second, which from the 50th, 100th, 150th, 200th, and 250th frames are shown here.}
    \label{fig:proto}
\end{figure}

One of the advantages of the SSR-PR approach is a full wavefront reconstruction of dynamic scenes. We recorded the movement of a  single-celled eukaryote, in real time for 10 seconds, resulting in a set of diffraction patterns containing 287 frames. Then the frames are post-processed using the proposed method. After the reconstruction of the whole image set, a Spatiotemporal Video Filtering\cite{maggioni2014joint} (RF3D) was applied on the reconstructed video footage to remove the random and fixed-Pattern noises. It is a modification of BM3D, therefore following a similar sparsity-based filtering, but taking several consecutive image frames into account.

This result is a breakthrough, since, contrary to the existing phase retrieval based algorithms\cite{ryu2017subsampled,zhang2018lensfree} the SSR-PR is capable of computational super-resolved video reconstruction via phase retrieval with an adequately high frame rate. As the best of our knowledge, this is unique of its kind. In Fig. \ref{fig:proto} a few frames are presented from the video. We can see the reconstructed amplitudes on the top row and the height maps calculated from the reconstructed phases on the bottom row. The full reconstructed video is attached to the \href{https://tuni-my.sharepoint.com/:v:/g/personal/peter_kocsis_tuni_fi/EZBhbTmu9zZJsSIIsmPLOk0BjVDk5I4YcN3V_sf0BAS6-Q?e=JTvZEW}{supplementary}.

%% file: Sections/6_Conclusion.tex
\section{Conclusion}
A novel approach and algorithm are proposed for single-exposure lensless super-resolution phase retrieval system using compensated wavefronts. 
% The compensation is based on extended prior knowledge, which for preliminary tests of the optical elements were initiated. 
This compensation aims to enhance the computational modeling of the optical image formation, to increase the correspondence with the physical system. It is based on preliminary tests of the optical elements, which might be treated as prior system calibration.
Notedly, two tests were made without the object, one with the binary modulation phase mask and one without. The latter one was used to to give an approximation of the carrying wavefront on the object plane, which is then enhanced by the proposed compensation. We analyzed the compensation and found that it is also correcting the noises appeared from the error of the optical elements and presumably the ill-posedness. The novel approximate of the object wavefront is used in the SSR-PR method, resulting in high-quality super-resolved reconstructions. The super-resolution is achieved by initial upsampling of the wavefronts. For upsampling, the default box kernel was replaced by Lanczos-3 kernel with stairstep interpolation, providing higher phase-correction. 
% The reconstructions are apodized by support constraints with ones and the field of interest is BM3D filtered.

We adjusted the simulations to fit to the physical system and reconstructed details 4$\times$ smaller than the sensor pixel size. Comparing with the previous SR-SPAR method, the difference is especially well marked in the case of small phase values. This corresponding well to the physical experiments, in which we were able to resolve 2 $\mu$m wide lines, as the smallest group of a calibrated test object, with 3.45 $\mu$m sensor pixel size. The biological application was demonstrated by high-detailed reconstruction of static Buccal Epithelial Cells without any preceding preparation. 
% The comparison with digital holographic imaging
We found that the SSR-PR approach provides much detailed reconstruction than the predecessor SR-SPAR, while significantly better phase correction than the traditional maskless phase retrieval. 
% We found that the SSR-PR approach provides significantly better phase correction than the traditional maskless phase retrieval.
A significant advantage of SSR-PR is the possibility to observe dynamic objects. We recorded the movement of a single-celled eukaryote and after post-processing we achieved an high frame-rate super-resolved video footage, which is an exceptional result.

% As for further work, we aim to further improve the approximations % correct the remaining phase-loss 
% and make the system more stable for in-field applications. 
A possibility for further work is to move into the multiwavelength direction with unique sensor filters to keep the single exposure property. This direction could also provide us color imaging capability. 
% For real in-field applications, the system has to be implemented in a portable prototype, which could be the next milestone in this research. 
Furthermore, the simple optical setup and appropriately small dimensions makes the system suitable for mobile imaging. Developing such integrated hardware could be the next milestone in this research.

\section*{Acknowledgements}
    
This research was supported by Jane and Aatos Erkko Foundation, Finland, and Finland Centennial Foundation: Computational Imaging without Lens (CIWIL) project and the Academy of Finland Flagship Programme, Photonics Research and Innovation (PREIN), decision 320166.